%% file: main.tex
\documentclass[
reprint,
superscriptaddress,
amsmath,amssymb,
aps,
prx,
floatfix,
longbibliography
]{revtex4-2}

\input{tikzmacros}
\usepackage{mathrsfs}
\usepackage{amssymb}
\usepackage{amstext}
\usepackage{amsmath}
\usepackage{mathtools}
\usepackage{color}
\usepackage[english]{babel}
\usepackage{hyperref}
\usepackage{graphicx}
\usepackage{bm}
\usepackage{float}
\usepackage{dcolumn}
\usepackage{appendix}
\usepackage{comment} 
\usepackage{tikz}
\usetikzlibrary{graphs}
\usepackage{ifthen}
\usepackage{pgfplots}
\usepackage{cleveref}
\usepackage[multiple]{footmisc}
\pgfplotsset{compat=1.18}

\def\be{\begin{eqnarray}}
\def\ee{\end{eqnarray}}
\def\nn{\nonumber}
\def\tt{|\tau\bar{\tau}\rangle}
\def\t1{|\tau {1}\rangle}
\def\1t{|{1}\bar{\tau}\rangle}
\def\oo{|{11}\rangle}
\def\ra{\rangle}
\def\fphi{\frac{1}{\phi}}
\def\fsqphi{\frac{1}{\sqrt{\phi}}}

\begin{document}

\title{Minimal Quantum Circuits for Simulating Fibonacci Anyons}
\author{Sary Bseiso}
\affiliation{Department of Physics, University of Illinois at Urbana-Champaign,
1110 West Green Street, Urbana, IL 61801, USA}
\author{Joel Pommerening}
\affiliation{Institute for Quantum Information,
RWTH Aachen University, D-52056 Aachen, Germany}
\author{Richard R. Allen}
\affiliation{Center for Theoretical Physics, Massachusetts Institute of Technology, Cambridge, MA 02139, USA}
\author{Steven H. Simon}
\affiliation{Rudolf Peierls Centre for Theoretical Physics, Parks Road, Oxford, OX1 3PU, UK}
\author{Layla Hormozi}
\email{hormozi@bnl.gov}
\affiliation{Computational Science Initiative, Brookhaven National Laboratory, Upton, NY 11973, USA}


\begin{abstract}
The Fibonacci topological order is the prime candidate for the realization of universal topological quantum computation. We devise minimal quantum circuits to demonstrate the non-Abelian nature of the doubled Fibonacci topological order, as realized in the Levin-Wen string net model.  Our circuits effectively initialize the ground state, create excitations, twist and braid them, all in the smallest lattices possible. We further design methods to determine the fusion amplitudes and braiding phases of multiple excitations by carrying out a single qubit measurement.  We show that the fusion channels of the doubled Fibonacci model can be detected using only three qubits, twisting phases can be measured using five, and braiding can be demonstrated using nine qubits. These designs provide the simplest possible settings for demonstrating the properties of Fibonacci anyons and can be used as realistic blueprints for implementation on many modern quantum architectures.

\end{abstract}

\maketitle
\section{Introduction}
The emergence of topological order in interacting quantum systems is one of richest phenomena in modern condensed matter physics~\cite{wen13}. The possibility of creating long-range entangled states~\cite{chen10} and quasi-particle excitations with non-Abelian statistics has opened the door to new vistas in fault-tolerant quantum information processing, both in the form of~(passively) error-tolerant topological quantum processors, and~(active) topological quantum error correcting codes (QECCs)~\cite{kitaev03,kitaev06,freedman01,freedman02-1,freedman02-2,nayak08,Simon_book}. For almost three decades, this two-way relationship has been a wellspring of creative work, from quantum algorithms with super-polynomial speedups~\cite{aharonov08, kuperberg15} to topological codes such as the surface code, which are among the most promising QECCs~\cite{fowler09, fowler12, terhal15}. 

In recent years, rapid progress in simulating topological phases has been achieved, and observation of Abelian statistics has been reported in a variety of platforms. This includes minimal implementations using optical qubits~\cite{han07, lu09} to intermediate-scale experiments in superconducting qubits~\cite{satzinger21}. More recently, progress has been reported on quantum simulation of certain non-Abelian phases, including dislocation-induced non-Abelian excitations in Abelian states~\cite{andersen22}, Majorana states~\cite{harle23}, as well as non-Abelian states, such as those with $D_4$ and $S_3$ topological order~\cite{iqbal23, steve23}.

Despite this progress, some of the most interesting topological phases, particularly those which can be used for universal topological quantum computation and quantum error correction, have remained out of reach. Fortunately, rapid advances in the science and engineering of controllable arrays of synthetic qubits, in various platforms, are increasingly enabling more advanced digital quantum simulation. Given efficient circuits which prepare and manipulate the ground states and quasiparticle excitations of topological phases of matter, it is now possible to directly probe their fascinating properties on a variety of nascent quantum hardwares.

\begin{figure*}[t]
\centering
\includegraphics[width = \textwidth]{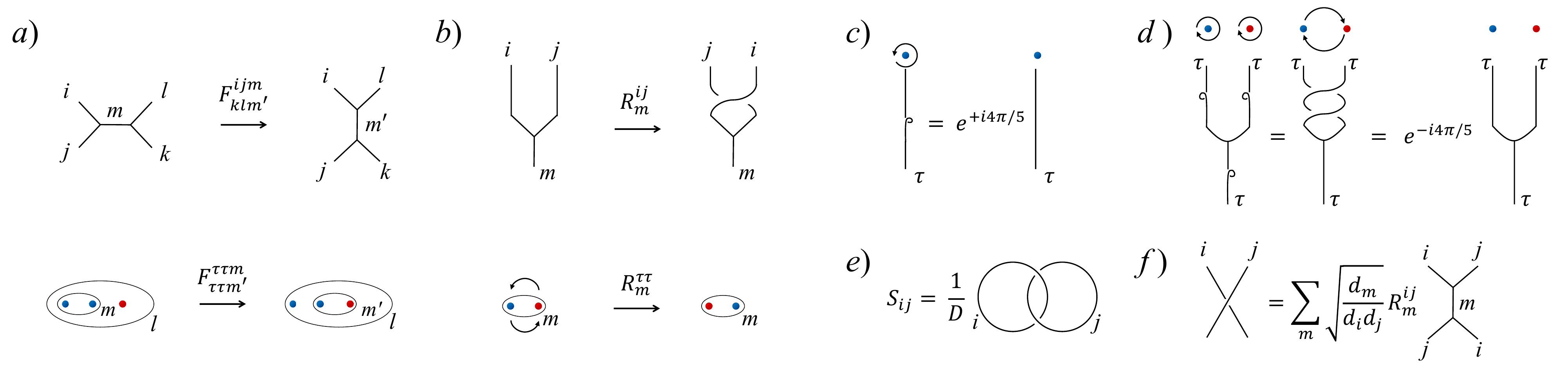}
\caption{(a) The $F$-move as a change of basis operator and its non-trivial matrix representation.  (b) The action of braid operator $R$ and its matrix representation. (c) The twist operator and (d) the relation between braiding and twisting.  (e) The topological $S$ matrix. (f) The relation between braiding and fusion. Here $d_m$ is the quantum dimension of anyon type $m$. This relation can be used to resolve string crossings.}
\label{fig:FRS}
\end{figure*}

In this article, we describe minimal quantum circuits which demonstrate the non-Abelian properties of Doubled Fibonacci (DFib) anyons, arising as excitations of Levin and Wen's string-net model~\cite{levinwen05}. Fibonacci topological order~\cite{preskillnotes}, is particularly interesting, since it is the simplest candidate for universal topological quantum computing, having braiding operations which are universal and can be implemented fault tolerantly~\cite{kkr,schotte20}. As suggested above, our approach is based on digital quantum simulation: applying a series of quantum gates which effectively project into the desired eigenstate of the string-net Hamiltonian.~\cite{bonesteel12, schotte20, liu22}. 

In constructing these circuits, we identify the minimal settings in which the intrinsically ``topological'' properties  of Fibonacci anyons (i.e., the building blocks for topological quantum computation) can be verified. In particular, each property is realized and measured using the fewest number of qubits possible, and with circuits that are sufficiently shallow to be accessible on virtually any modern quantum hardware. Though our constructions generalize to larger system sizes, asymptotic scaling is not the focus of this paper. Rather, we aim to build a specific set of primitive subroutines to enable further progress in the experimental understanding of topological phases of matter on programmable quantum devices.

This paper is organized as follows. In Sec.~\ref{sec:fibbasics}, we review the essential features of both the chiral Fibonacci and the achiral DFib models and establish the notation. In Sec.~\ref{sec:init}, we describe our approach to initialization and measurement of DFib anyons. In Sec.~\ref{sec:signs}, we discuss the details of constructing minimal circuits for demonstrating fusion, twisting and braiding of DFib anyons in different settings. Our circuits use, three, five and nine qubits, respectively --- the smallest number possible in each case. Section~\ref{sec:sum} concludes and provides an outlook.

\section{The Fibonacci Topological Order}
\label{sec:fibbasics}
In this section, we first briefly review the (chiral) Fibonacci anyon model ($\mathrm{Fib}$), then describe the basics of the Levin-Wen string-net model~\cite{levinwen05} that realizes the (achiral) DFib topological order.  

\subsection{The Chiral Fibonacci Model}
The Fibonacci anyon model~\cite{preskillnotes,Simon_book} can be described by a set of anyon types, here represented by $1$ and $\tau$, where the former represents the trivial (vacuum) state and the latter the only non-trivial anyon type, carrying a quantum number known as \textit{topological charge}. Similar to other quantum numbers (e.g., spin or electric charge), there are rules for combining topological charge. In the case of Fibonacci anyons, these rules are: 
\be 
\tau \times 1 = \tau,\\\nn
1 \times 1 = 1,\\\nn
\tau \times \tau = 1 + \tau.
\label{eq:FibFusion}
\ee
The first two equations mean that combining (fusing) any object of charge $1$ or $\tau$ with the vacuum state $1$ will not affect the charge of that object. The third equation means that combining  the topological charges of two objects with charge $\tau$ will result in either an object with trivial charge 1 or an object with charge $\tau$.

An important consequence of the Fibonacci fusion rule is that the Hilbert space of $n$ Fibonacci anyons, each of charge $\tau$, has a dimension which grows according to the Fibonacci sequence~(hence the name). In the asymptotic limit, the dimensionality of the Hilbert space grows as $\phi^n$, where
$$
\phi = \frac{\sqrt{5} + 1}{2}
$$ 
is the golden ratio. Since $\phi$ sets the growth rate of Hilbert space in this model, it is referred to as the \textit{quantum dimension} of Fibonacci anyons, $d_\tau = \phi$. The quantum dimension of the vacuum particle is unity $d_{1} = 1$.

The order of anyon fusion defines a choice of basis for the Hilbert space. For example, given three anyons in a row, we can choose to fuse the first two, then add the third one. Equivalently, we can choose to first fuse the last two, then add the first one. These two choices define two different bases for the three-dimensional Hilbert space of three anyons. These two bases can be mapped to each other by a unitary operation known as $F$ operator, or $F$-move~\cite{bonesteel12}, defined pictorially in Fig.~\ref{fig:FRS}(a). In its simplest non-trivial form, the $F$  operator has the following matrix representation:
\be
\label{eq:Fmatrix}
F \equiv F^{\tau\tau m}_{\tau\tau m'} = \begin{pmatrix}
\frac{1}{\phi} &  \frac{1}{\sqrt{\phi}} \\
\frac{1}{\sqrt{\phi}}  & -\frac{1}{\phi} \\
\end{pmatrix},
\ee
where $m,m' \in \{1, \tau\}$.

In the quantum computing community, Fibonacci anyons are best known for the universality of their braiding: their braid operators are generators of $\mathrm{SU}(2)$, and hence can be used to approximate all single-qubit gates with arbitrary accuracy~\cite{freedman02-1, freedman02-2}. In general, the braid operator $R^{ij}_m$ describes the phases resulting from a counterclockwise exchange of two anyons with topological charges $i$ and $j$, such that the total charge of the system is $m$, as depicted in Fig.~\ref{fig:FRS}(b). The matrix representation of the non-trivial  braiding of two Fibonacci anyons with total charge $m$ is as follows:
\be
R 
\equiv R^{\tau \tau}_m 
=  \begin{pmatrix}
e^{-4\pi i/5} &  0 \\
0  & e^{+3\pi i/5} \\
\end{pmatrix}.
\ee
The diagonal nature of this matrix indicates that braiding two anyons does not change their total charge. The first and second diagonal elements correspond to the phase acquired by braiding two Fibonacci anyons with total charges $1$ and $\tau$, respectively. These are the braiding phases for so-called ``right-handed'' Fibonacci anyons; we can define a separate left-handed Fibonacci model whose braiding phases are the complex conjugate of those above. 

\begin{figure}[t!]
\centering
\includegraphics[width = .49\textwidth]{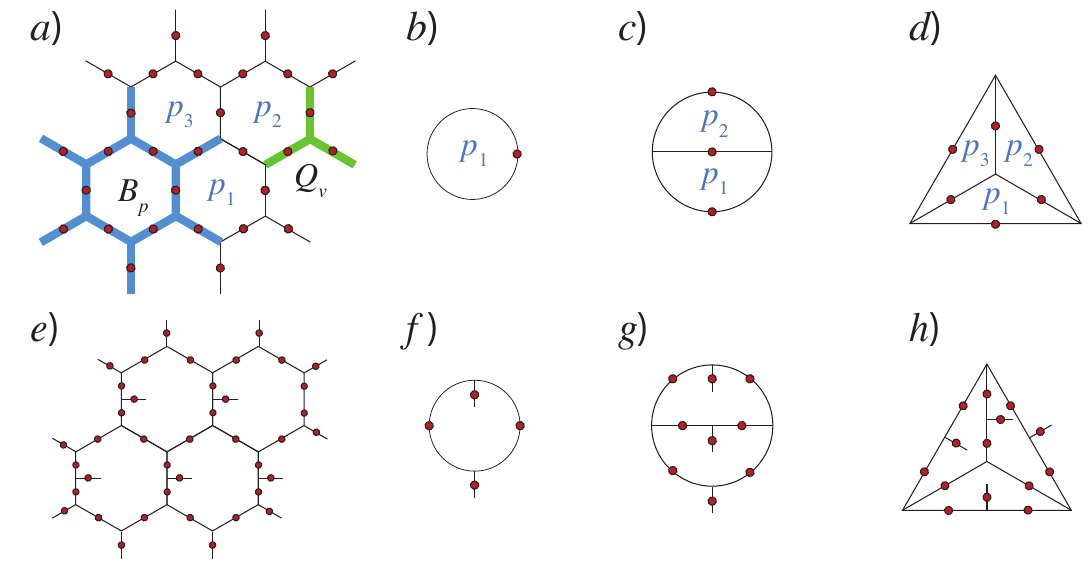}
\caption{Basic 2D trivalent lattices that are used in this work, all with spherical boundary conditions. Qubits sit on the edges and are depicted by red circles. 
(a) A patch of four plaquettes out of a larger hexagonal lattice. Here we also show the plaquette and vertex operators $B_p$ and $Q_v$, respectively. In this lattice, $B_p$ acts on the 12 edges surrounding plaquette $p$ (shaded in blue) and $Q_v$ acts on the three edges surrounding the vertex $v$ (shaded in green). (b) The simplest lattice, consisting of one edge. It can also be thought of as a tadpole with a trivial tail. Assuming spherical boundary conditions, this edge is shared between two plaquettes, $p_1$, the plaquette shown, and $p_2$ the outside of the plaquette which represents the remainder of the sphere (not shown). (c) Three plaquettes on the $\Theta$ lattice. Plaquette $p_3$ is the outside of the diagram and includes the remainder of the  sphere. (d) Four plaquettes on a tetrahedron, where again the fourth plaquette is the remainder of the sphere. (e-h) Counterparts of (a-d) on a tailed lattice, where two additional qubits are added to each plaquette. We call (f) a ``generalized'' tadpole.}
\label{fig:lattice}
\end{figure}

Closely related to braiding is the concept of twisting, or a counterclockwise rotation of  an anyon around itself by $2\pi$, as shown in  Fig.~\ref{fig:FRS}(c). The action of twisting is encoded in the  topological twist operator, or topological spin, $\theta_i$, which is related to the braid operator as follows, 
\be 
\theta_i = (R^{i i}_1)^{-1}.
\ee
It follows that braiding and twisting are related to each other as,
\be
\label{eq:braidtwist}
(R_m^{ij})^2 = {\theta}_m{\theta}_i^{-1}{\theta}_j^{-1},
\ee
provided that 
$i$ and $j$   fuse to $m$.
Here  $i,j,m\in\{1, \tau\}$, ${\theta}_1 = 1$ and ${\theta}_\tau = e^{+4\pi i/5}$. A visual depiction of this relation is shown in Fig.~\ref{fig:FRS}(d).

Another element of the Fibonacci anyon model that will be relevant to us is the modular $S$-matrix, which has the form,
\be
S 
= \frac{1}{D}\begin{pmatrix}
1 &  \phi \\
\phi & -1 \\
\end{pmatrix}.
\label{eq:Smatrix}
\ee
Here $D = \sqrt{1 + \phi^2}$ is the so-called \textit{total quantum dimension} of the Fibonacci model. This operator describes the process in which two pairs of $\tau$ particles are created out of vacuum, then two anyons from different pairs exchange twice (one making a full wrap around the other), and finally each re-annihilates with its respective partner, forming the so-called ``Hopf link.’’ A visual description of this operator is also given in Fig.~\ref{fig:FRS}(e).

Finally, braiding and fusion rules can be used to derive relations to resolve string crossings. A version of this relation, which will be used extensively in the next section, is given in Fig.~\ref{fig:FRS}(f).

\begin{figure}[t]
\centering
\includegraphics[width = .495\textwidth]{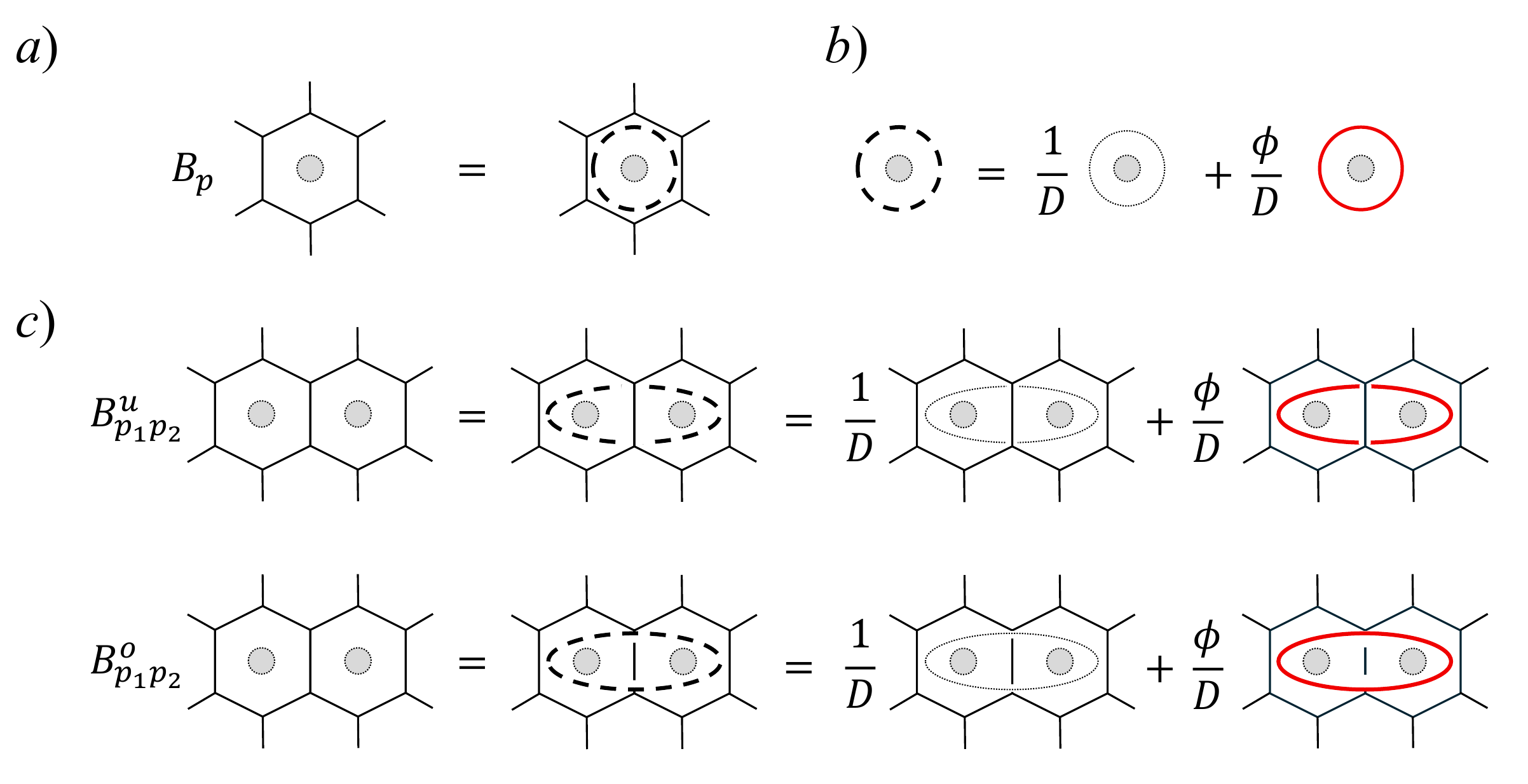}
\caption{(a)~The effect of $B_p$ operator is to add a vacuum loop to the plaquette $p$, which can then be absorbed through the edges. The gray core indicates a punctured lattice, preventing the  loops around it from contracting. (b) The vacuum loop (dashed line) consists of a trivial loop with charge $1$ (dotted line), and a $\tau$ loop (solid red line). (c)~The analogous two-plaquette operators $B^o_{p_1 p_2}$ and $B^u_{p_1 p_2}$, acting on two plaquettes by inserting a vacuum loop over and under the dividing edge, respectively.}
\label{fig:Bp}
\end{figure}

There is no known exactly solvable model for the realization of the chiral Fibonacci model. However, Levin and Wen's string-net construction provides a framework for the doubled (achiral) realization of all types of topological order, including DFib. This model can be thought of as two copies of the Fibonacci model with opposite chiralities, where fusion and 
braiding rules apply independently to the right-handed $\tau$ and the left-handed $\bar\tau$ excitations. The resulting DFib topological order, $\mathrm{Fib} \otimes \overline{\mathrm{Fib}}$,  has four distinct anyon types: $\{1,\tau\} \otimes \{1, \bar{\tau}\} = \{ 11, \tau1, 1\bar{\tau}, \tau\bar{\tau}\} $. Our realization of the DFib topological order is based on Levin and Wen's original~\cite{levinwen05} and extended string-net models~\cite{feng15, hu18, schotte20}. In what follows we will briefly describe the properties of this model. 

\subsection{String-net Realization of the DFib Model}

\begin{figure*}[t!]
\centering
\includegraphics[width = \textwidth]{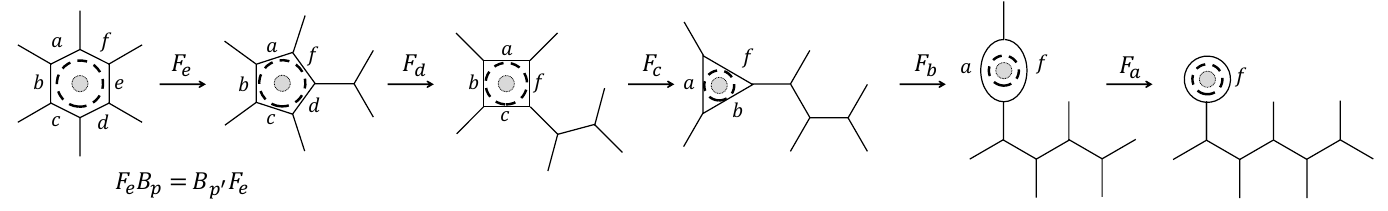}
\caption{The plaquette operator $B_p$ and the $F$-move commute with each other: The process of first measuring the charge of a plaquette $p$, then reducing its size by applying an $F$-move to the edge $e$ provides the same result as first applying the $F$-move and then measuring the charge of the new plaquette $p'$. As a consequence, if the plaquette $p$ is in the ground state, the reduced plaquette $p'$ is also in the ground state. By continuing the process we can reduce a generic $n$-sided plaquette to a tadpole~\cite{koenig09, bonesteel12}. Here the  dashed loop represents the vacuum loop, which can be absorbed through the edges at each step. The gray dot in the middle of the plaquette indicates that the plaquette is punctured, thus the vacuum loop cannot be contracted. 
}
\label{Fig:FBp}
\end{figure*}

The Levin-Wen (LW) string-net model~\cite{levinwen05} is defined by a set of string types on the edges of a 2D trivalent lattice and a set of self-consistent rules for combining (fusing) different string types. These fusion rules  define the string-net Hilbert space. A family of exactly solvable Hamiltonians can be introduced in this Hilbert space, which realize all achiral topological phases, and can be thought of as a generalization of Kitaev's toric code~\footnote{Technically,  a string net model can be defined for any ``unitary fusion category'' $\cal C$, resulting in an  anyon theory which is the so-called Drinfeld center $Z(\cal C)$.}. These Hamiltonians have the following form, 
\be
H = -\sum_v Q_v -\sum_p B_p,
\ee
where $Q_v$ and $B_p$ are mutually commuting vertex and plaquette projection operators. The former acts on the three edges surrounding the vertex~$v$, while the latter acts on the edges surrounding the plaquette~$p$. Fig.~\ref{fig:lattice}(a) shows these operators for a hexagonal lattice. The system is in its ground state when both the vertex and plaquette operators are\textit{satisfied} with eigenvalue~$+1$.   We say an operator is {\it violated} if it has eigenvalue 0 rather than $+1$, and this corresponds to an excited state of the system.

The operators $Q_v$ and $B_p$ act on the space of ``string types,'' which can be encoded as multi-level spins, or \textit{qudits}, in general. In this picture, the vertex operator enforces certain branching rules, consistent with the fusion rules of the strings. In particular, the branching rules decree that $Q_v$ is always violated if a single non-trivial string ends at a vertex (an open string)~\footnote{This means a single anyon cannot be created out of vacuum.}. Thus, non-trivial string types that satisfy $Q_v$ always form closed configurations (loops or nets). We will refer to the space of states for which all $Q_v$'s  are satisfied as the ``string-net space.''

The plaquette operator $B_p$ effectively measures the topological charge of plaquette $p$ and is satisfied when the plaquette contains trivial topological charge. This operator, which also controls the dynamics of string-nets, imposes further restrictions on the string-net space. Together with $Q_v$ these operators favor a certain superposition of loops and nets, which defines the ground state of the string-net model. In the original Levin-Wen model, various excitations result from different combinations of $Q_v$ and $B_p$ violations.

The DFib topological order~\cite{Fidkowski2009,Simon_book} is a specific instance of the LW string-net model, with two string types, corresponding to $\{1,\tau\}$ in the chiral Fibonacci phase. Here we identify the string types $1$ and $\tau$ with qubit states $|0\ra$ and $|1\ra$, respectively. We will sometimes refer to an edge with string type $\tau$, or equivalently qubit state $|1\ra$, as an \textit{activated} edge. The vertex operators, $Q_v$, essentially enforce branching rules  that are consistent with the fusion rules of the Fibonacci topological order Eq.~\eqref{eq:FibFusion}. As a result, $Q_v$ is only violated when a single incoming edge at vertex $v$ is activated.

The plaquette operator in this case can be written as,
\be 
B_p = \frac{1}{D^2}(1 + \phi B_p^\tau),
\ee
where the operator $B_p^\tau$ effectively adds a loop of string type $\tau$ to the plaquette $p$. In fact, $B_p$ can be thought as of an operator that adds what is known as a \textit{vacuum loop}, a weighted sum of vacuum and a $\tau$ loop, to the plaquette, which can then be absorbed into the surrounding edges using a series of $F$-moves~\footnote{For a detailed description of the process of absorbing strings through lattice edges with a series of $F$-moves see the original work in~\cite{levinwen05}}. Fig.~\ref{fig:Bp} depicts  the vacuum loop and the effect of the $B_p$ operator on plaquette $p$.

As was mentioned above, the DFib model  has four  anyon types, $\{ 11, \tau1, 1\bar{\tau}, \tau\bar{\tau}\}$. Here $11$ indicates the ground state, corresponding to no topological charge, which results from satisfying all the $Q_v$ and $B_p$ operators. In the original string-net model, the excitations $\tau\bar{\tau}$, $\tau1$ and $1\bar\tau$ result from combinations of $Q_v$ and $B_p$ violations. 

 The $\tau\bar{\tau}$ excitation consists of four components: $\tau\bar{\tau}_{11}, \tau\bar{\tau}_{1\tau}, \tau\bar{\tau}_{\tau1}$ and $\tau\bar{\tau}_{\tau\tau}$, all of which have the same topological charge~\cite{levinwen05, kkr, schotte20}. Out of these, the $\tau\bar{\tau}_{11}$ component can be realized just by violating $B_p$, thus it can be realized in the original LW model without leaving the string-net space. This excitation is essentially achiral since the phases resulting from braiding $\tau$  and $\bar{\tau}$ generally cancel each other. However, due to its multi-channel fusion properties, a restricted set of braiding phases can be observed for these anyons. We will discuss this in more detail in Sec.~\ref{sec:signs}.

To demonstrate the full braiding phases of the DFib model, we need to consider the chiral excitations $\tau 1$ or $1 {\bar\tau}$. These excitations possess braiding statistics that are essentially equivalent to that of chiral Fibonacci anyons (right- and left-handed, respectively) and can therefore realize universal quantum computation~\cite{freedman02-2,kkr}.

To realize these excitations without violating $Q_v$, we use the extended LW model, which is defined on a \textit{tailed} lattice with two extra edges added to each plaquette~\cite{feng15, hu18, schotte20}, as shown in Fig.~\ref{fig:lattice}(e-h). The addition of these tails will allow us to correct $Q_v$ violations while preserving the excitations, thus effectively permitting the creation of chiral excitations within the string-net space.

\section{Initialization and Charge Measurement} 
\label{sec:init}
%

\begin{figure*}[t]
\centering
\includegraphics[width = \textwidth]{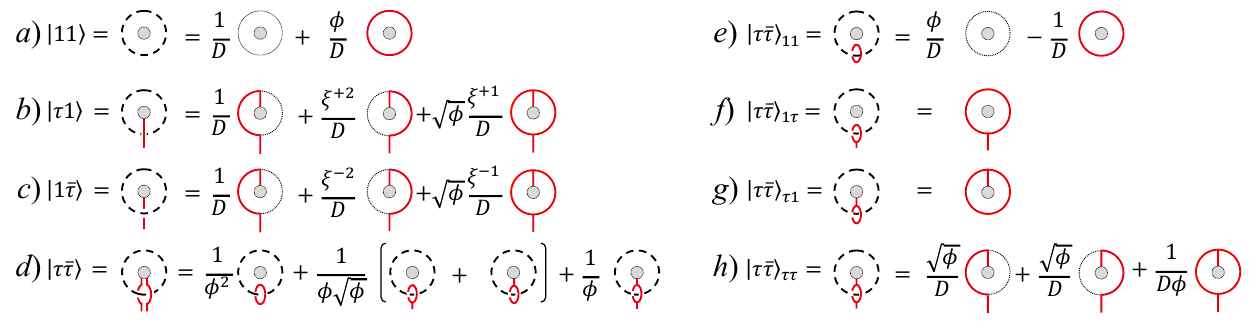}
\caption{Excitation strings and vacuum loops. Here $\xi = e^{3\pi i/5}$, and the dashed lines represent vacuum loops, crossed by red $\tau$ strings in various ways, giving rise to generalized tadpole representations of the four anyon types of the DFib model. (a) The ground state $\oo$ corresponds to a vacuum loop without any string crossings; (b) $\t1$ resulting from a $\tau$ string crossing ``over'' the vacuum loop; (c)  $\1t$  resulting from a $\tau$ string crossing ``under'' a vacuum loop and (d) the $\tt$ sector where two $\tau$ strings cross over and under the vacuum loop. The end points of these  strings can combine in four different ways, resulting in the four components of $\tt$, shown in (e-h). }
\label{fig:basistadpoles} 
\end{figure*}

The string-net space is defined by satisfying the vertex operator $Q_v$ on every vertex $v$.  We can project to this space by measuring this operator at every vertex and then correcting where an error is detected. 
Relatively simple circuits for measuring the vertex operator and correcting the affected vertices have been introduced~\cite{bonesteel12, schotte20}. Here, we do not explicitly show the vertex measurement/correction circuits, but we assume all $Q_v$ operators are satisfied before applying our circuits.

Projection to the ground state requires further satisfying $B_p$ operators at every plaquette, thus selecting a particular superposition of string-nets, which is a highly entangled state. Since the plaquette operators are projectors, we can ensure they are satisfied by a projective measurement, though the measurement itself is very nontrivial, as we shall now see.

We follow the method introduced in~\cite{koenig09, bonesteel12}, which is inspired by the idea of entanglement renormalization~\cite{vidal07}. This method is based on the observation that the $F$-move acting on a given edge $e$, which effectively redraws the lattice at that edge, and the operator $B_p$, which effectively measures the  charge of plaquette $p$,  ``commute" with each other in the following sense: 
\be
F_e B_p = B_{p'} F_e.
\label{eq:FBp}
\ee
This means that if we measure the charge of an $n$-sided plaquette $p$ with operator $B_p$, then apply an $F$-move to an adjacent edge $e$ to reduce the size of the plaquette, the result is the same as applying the $F$-move first, followed by $B_{p'}$ applied to the reduced-size $(n-1)$-sided plaquette $p'$ (see Fig.~\ref{Fig:FBp}). An important consequence of this observation is that $F$-moves can be used to locally modify the lattice while preserving the ground state. 

Following~\cite{bonesteel12}, we use this fact to both initialize the plaquettes and also to measure their charge. The basic idea is that by repeatedly applying $F$-moves to the edges of an $n$-sided plaquette, we can reduce it to a single tadpole: a two-edge graph defined by a a closed edge forming the ``head''  and a connecting edge forming the ``tail'': 
$$
\tadpole{1}{1}{1}.
$$
We can use the above rule to reduce any lattice to a series of tadpoles, which can be easily initialized to the $11$ ($B_p = 1$) state using single qubit gates , then reverse the $F$-moves to return to the initial lattice, now in its ground state~\cite{bonesteel12}. Similarly, to measure the charge of a plaquette, we can reduce it to a tadpole with the same procedure, then carry out single qubit measurements of the head and tail edges to determine the charge of the original plaquette. In what follows, we discuss how this notion can be  generalized to initialize and measure the charges of other states with anyon types $\tau1, \ 1\bar{\tau}$ and $\tau\bar{\tau}$. We also generalize this notion to devise a procedure for measuring the total charges of multiple plaquettes. 

\subsection{Excitations}  Not only do $F$-moves preserve plaquettes satisfying $B_p = 1$, but the statement that $F$ and $B_p$ operators commute (Eq.~\eqref{eq:FBp}) guarantees that this holds for all eigenstates of the LW Hamiltonian. We can use this fact to both create and measure the entire DFib spectrum. 

As noted earlier, in the original string-net model with no tails, the only other anyon type that can be initialized with a tadpole, without leaving the string-net basis, is $\tau\bar{\tau}$. To realize the chiral excitations $\tau1$ and $1\bar{\tau}$ without violating the vertex operator, we use the tailed lattice~\cite{feng15, hu18, schotte20} in which each plaquette is equipped with two extra edges realizing a tail (see~Fig.~\ref{fig:lattice}).  Similar to the original tail-less plaquette, we can reduce the tailed plaquette using a series of $F$-moves. The result is a \textit{generalized tadpole} where an  extra \textit{inward} tail is added inside the head of the tadpole:
\be
\nn
\gentadpole{1}{1}{1}{1},
\ee
(see also Fig.~\ref{fig:lattice}(f)). This generalized tadpole is minimally represented by four qubits and  can be used to initialize all four anyon types of the DFib model, including the four components of the $\tau\bar\tau$ anyon.

Just as the ground state can be created by initializing tadpoles with a vacuum loop, excitations can be created by superimposing $\tau$ strings to the vacuum loops~\cite{levinwen05, kkr}. This process is shown in Fig.~\ref{fig:basistadpoles}. If a string passes over a vacuum loop, it results in  a generalized tadpole which generates the $\tau 1$ excitation. The crossings between the $\tau$ strings and the vacuum loops can be resolved by using the relation given in~Fig.~\ref{fig:FRS}(f). 

Likewise, if a string passes under a vacuum loop, it leads to a $1\bar\tau$ generating (generalized) tadpole. If two strings pass over and under the vacuum loop, the result is a $\tau \bar\tau$ generating tadpole. Note that the end points of two strings passing over and under the vacuum loop can fuse in four possible ways, resulting in the four components of the $\tau\bar\tau$ anyon. In the tailed lattice, these $\tau$ strings must end on the added tails to avoid $Q_v$ violations. These are the same tails that show up in the generalized tadpoles that generate the DFib spectrum, a summary of which is also given below,
\be
\label{11}
\oo =  \frac{1}{D}\tps{0}{0}{0}{0} +  \frac{\phi}{D}\tps{1}{1}{0}{0}\\
\label{t1}
\t1 =  \frac{1}{D}\tps{0}{1}{1}{1} +  \frac{\xi^{2}}{D}\tps{1}{0}{1}{1} +   \sqrt{\phi} \: \frac{\xi}{D}  \tps{1}{1}{1}{1}\\
\label{1t}
\1t =  \frac{1}{D}\tps{0}{1}{1}{1} +  \frac{\xi^{-2}}{D}\tps{1}{0}{1}{1} +   \sqrt{\phi} \: \frac{\xi^{-1}}{D}  \tps{1}{1}{1}{1}\\
\label{tt11}
\tt_{11} =  \frac{\phi}{D} \tps{0}{0}{0}{0} -  \frac{1}{D}\tps{1}{1}{0}{0}\\
\label{ttt1}
\tt_{1\tau} =  \tps{1}{1}{1}{0}\\
\label{tttt}
\tt_{\tau1} =  \tps{1}{1}{0}{1} \\
\label{tt1t}
\tt_{\tau\tau} =  \frac{\sqrt{\phi}}{D}\tps{0}{1}{1}{1} +  \frac{\sqrt{\phi}}{D}\tps{1}{0}{1}{1} +   \frac{1}{D \phi}  \tps{1}{1}{1}{1}.
\ee
Here the red (shaded) lines indicate  strings of type $\tau$ (or qubit state $|1\ra$), while dotted  lines depict the type $1$ strings (qubit state $|0\ra$), and $\xi = e^{3\pi i/5}$. 

\begin{figure*}[t]
\centering
\includegraphics[width = .8\textwidth]{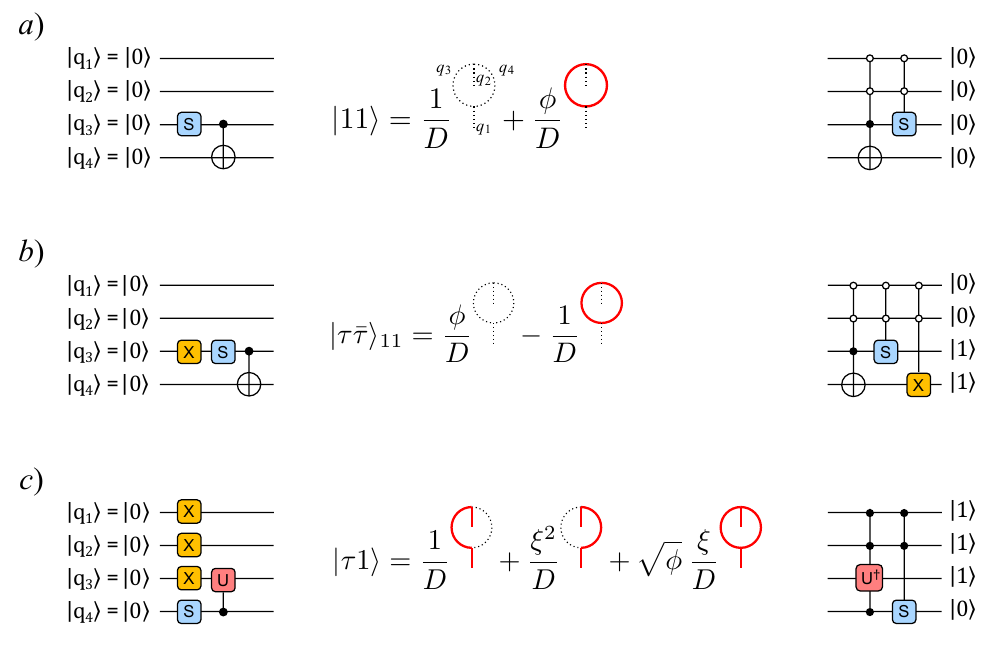}
\caption{Initialization and simplification circuits for (a) $\oo$, (b) $\tt_{11}$ and (c) $\t1$ excitations. 
In the $|q_1q_2q_3q_4\ra$ basis, these states have the following forms: $\oo = (|00\ra + \phi |11\ra)|00\ra/D$, 
$\tt_{11} = (\phi|00\ra - |11\ra)|00\ra/D$, and 
$\t1 =(|10\ra + \xi^{2}|01\ra + \xi \sqrt{\phi}|11\ra)|11\ra/D$, where 
$\xi = e^{i3\pi/5}$. Here $X$ is the NOT gate, and the $S$ and $U$ gates are defined in Eqs.~\eqref{eq:Smatrix} and \eqref{eq:Umatrix}, respectively. 
The initialization circuits are shown on the left and simplification circuits on the right. After simplifications $\oo$, $\tt$ and $\t1$ will be mapped to $|0000\ra$, $|0011\ra$ and $|1110\ra$, respectively, in the qubit basis.
}
\label{fig:init} 
\end{figure*}

Thus, to create a pair of excitations of any kind, we first reduce the lattice to a series of generalized tadpoles using  $F$-moves. Then, depending on the desired excitation, we initialize them according the rules given in Eqs.~\eqref{11}-\eqref{tttt}. Finally we return to the original lattice by reversing the $F$-moves.  The circuits for both initialization and measurement of these generalized tadpoles are shown in Fig.~\ref{fig:init}. The $S$ gate used in the figure is the modular $S$-matrix (Eq.~\eqref{eq:Smatrix}) while the $U$ gate is defined as,
\be
U = \begin{pmatrix}
\frac{\xi^{-1}}{\sqrt{\phi}} & \frac{\xi^{2}}{\phi} \\
-\frac{\xi^{-2}}{\phi} & \frac{\xi}{\sqrt{\phi}} \\
\end{pmatrix}.
\label{eq:Umatrix}
\ee

Finally, note that given the spherical boundary conditions, the excitations are always created in pairs --- all of the quantum numbers of all the excitations in the entire system must fuse to the identity.   Thus we cannot create a single plaquette excitation from the ground state without also creating its partner.   In Fig.~\ref{fig:basistadpoles} (and also in Eqs.~\eqref{11}-\eqref{tttt}) for simplicity, we show only half of the process, where a string is superimposed on a single vacuum loop inside the head of a (generalized) tadpole. In general, we need to form the tadpoles such that plaquettes that share a pair of excitations will be transformed to tadpoles that share tails, which can then be initialized together. We will discuss this in more detail the following section.

\begin{figure*}[t]
\centering
\includegraphics[width = .8\textwidth]{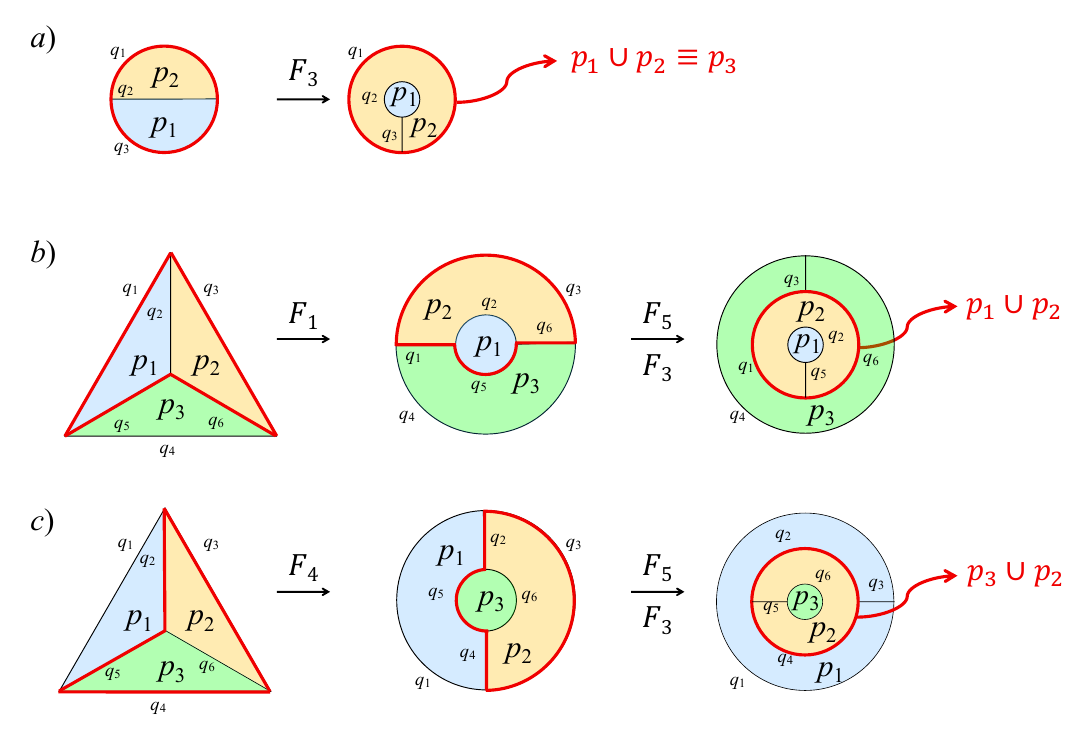}
\caption{Change of basis from lattice basis to concentric tadpoles used for initialization and measurement. The background colors keep track of the charge content of each area. The red lines mark the boundary of the two plaquettes that are transformed into a generalized tadpole. (a) Starting from the $\Theta$ lattice, we can change basis to concentric tadpoles by applying an $F$-move to one of the outer edges, in this case $q_1$. In the resulting lattice, the outer tadpole contains the information on the total charge of plaquettes $p_1\cup p_2$ in the original lattice, emphasized by the red boundary in both lattices. (b) Change of basis from a tetrahedron lattice to concentric tadpoles such that the total charge of plaquettes $p_1\cup p_2$ is enclosed by the middle (generalized) tadpole. This change of basis can be carried out by three $F$-moves as shown. (c) Another change of basis for the tetrahedron lattice such that now the middle tadpole in the concentric tadpole basis contains the total charge of $p_2\cup p_3$. This change of basis can be carried out with another set of $F$-moves. }
\label{fig:concen}
\end{figure*}

\subsection{Multiple Plaquettes} 
In the extended LW model defined on lattices with tails where excitations are created in the absence of $Q_v$ violations, they are essentially created on plaquettes. Thus, to determine the fusion amplitudes resulting from combing several excitations, we need a method to measure the total charge of multiple plaquettes. We discuss this problem from two different perspectives, first by building projective operators that determine the charge of two (or more) plaquettes, then designing quantum circuits which allow us to measure the total charge of multiple plaquettes with a computational basis measurement.

As was noted in Sec.~\ref{sec:fibbasics}, the action of the plaquette operator $B_p$ is to determine the charge content of a plaquette (whether it is trivial or not) by inserting a vacuum loop inside the plaquette, and absorbing it through the edges by a sequence of $F$-moves~(see Fig.~\ref{fig:Bp}). This idea  can be generalized to an operator that determines the charge content of two neighboring plaquettes by inserting  a vacuum loop inside the two plaquettes. However, in this case, there is a choice of whether to insert the loop \textit{over} or \textit{under} the dividing edge. It turns out that these two options lead to two different operators, which we call $B^o_{p_i p_j}$ and $B^u_{p_i p_j}$, where the superscripts $o$ and $u$ refer to \textit{over} and \textit{under} respectively. A visualization of these operators is depicted in Fig.~\ref{fig:Bp}.

Satisfying the operator $B^o_{p_i p_j}$ indicates that the total charge of the right-handed anyons $\t1$ in the two plaquettes $p_i$ and $p_j$ is trivial. Likewise, satisfying  $B^u_{p_i p_j}$ implies that the total charge of left-handed anyons $\1t$ on the corresponding plaquettes is trivial. Satisfying both of these the operators $B^o_{p_i p_j}$ and $B^u_{p_i p_j}$  indicates that the total charge content of the two plaquettes is trivial. Note that this can be true even if individual plaquette operators $B_{p_i}$ and $B_{p_j}$ are not satisfied. This is analogous to the case where the collective charge of two $\tau$ anyons can be trivial.

Similar to single plaquette operators $B_p$, the double plaquette operators defined above are also projection operators, hence  cannot be implemented directly as a  unitary quantum circuit but instead must be implemented as a (rather complicated) projective measurement. Here we design quantum circuits to directly measure the charge of multiple plaquettes. Our approach is to generalize the single plaquette strategy --- initialization and charge measurement by first reducing to a  tadpole --- to a charge measurement by reducing multiple plaquettes to a series of \textit{concentric circles} with each circle tied to the next by a single edge, resembling a series of concentric generalized tadpoles. Examples of this process for small lattices are shown in Fig.~\ref{fig:concen}. 

The key idea is that, through a series of $F$-moves, we reduce one of the plaquettes in the original lattice to the central tadpole in the concentric circles basis. The next-outermost (generalized) tadpole then contains the total charges of the original plaquette \textit{and} another plaquette, which is reduced to an annulus surrounding the central tadpole. Similarly, with more plaquettes, the outer tadpoles hold the charge of all the original plaquettes which are subsequently enclosed by those tadpoles. Since the system should be thought of as living on the surface of a sphere, the outermost ring can be viewed as a tadpole as much as the innermost ring.

By choosing different series of $F$-moves, we can select different concentric tadpole bases that correspond to different combinations of plaquettes and use it as a basis for both initialization and measurement of combined charges. As an example, Fig.~\ref{fig:concen}(b) and (c) depict two different choices of basis, suitable for measuring the charges of plaquettes $p_1 \cup p_2$ or $p_3 \cup p_2$.~\footnote{Note that, in general, different sequences of $F$-moves can be used to transform a lattice into different concentric tadpoles. The  number of these choices increases in the case of tailed lattices and sometimes there are multiple sequences of $F$-moves that lead to seemingly similar concentric tadpoles. However, in general different choices can lead to additional twists and sometimes braids. Thus, caution needs to be exercised when selecting the $F$-move sequence to avoid unwanted phases. }

The equivalence of this approach to the projective operators above can be understood by recalling the action of the two-plaquette operator $B_{p_i p_j}$, which inserts a vacuum loop inside the two plaquettes. By keeping track of the $F$-moves which modify the lattice into concentric tadpoles, we can see that this vacuum loop effectively covers the area around two concentric tadpoles.

\section{Signatures of non-Abelian Anyons in minimal lattices}
\label{sec:signs}
\begin{figure}[t]
\centering
\includegraphics[width = .49\textwidth]{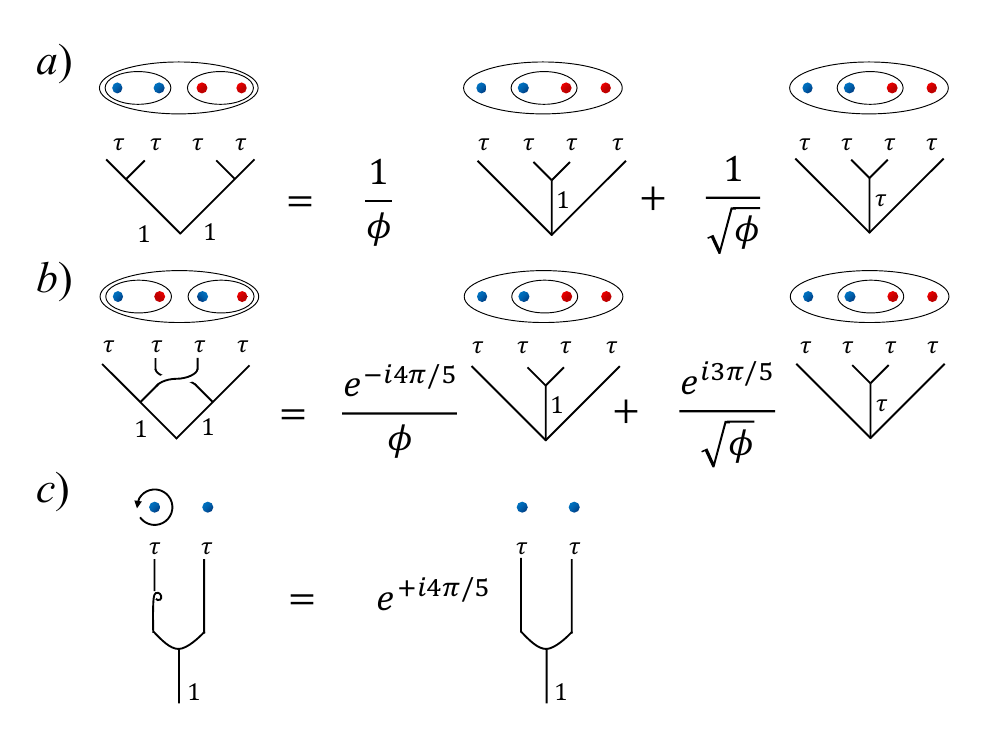}
\caption{Fusion, braiding, and twisting of chiral Fibonacci $\tau$ anyons. (a) Creation of two pairs of anyons from vacuum and cross-fusion. (b) Braiding of anyons and the corresponding phases assigned to each fusion channel. (c) Twisting a single anyon. }
\label{fig:fbt}
\end{figure}

In this section, we provide an overview of the signatures of the DFib model that we investigate in various settings, postponing the details to the following sub-sections. The basic properties that we are interested in are fusion, braiding and twisting of various excitations within the DFib model using the minimum number of qubits. Fig.~\ref{fig:fbt} depicts these operations for (chiral) Fibonacci anyons. 

We restrict our study to lattices with spherical boundary conditions. The simplest lattices that can be defined on a sphere are shown in Fig.~\ref{fig:lattice}. The smallest possible lattice  consists of  two plaquettes each forming a hemisphere with a single common edge forming the meridian (see Fig.~\ref{fig:lattice}(b)). This two-plaquette lattice can be initialized into ground state or host a pair of $\tau {\bar \tau} = \tau {\bar \tau}_{11}$ excitations, with total charge $11$. The former can be realized by by applying an $S$ gate to the single qubit on the edge initialized at $|0\ra$, while the latter by applying an $S$ gate to the qubit on the edge initialized at $|1\ra$, as can be seen in~Fig.~\ref{fig:init}.

The smallest non-trivial lattice for our purposes consists of three plaquettes, separated by three edges that meet at the two poles of the sphere. A 2D depiction of this lattice, which resembles the Greek letter $\Theta$, is shown in Fig.~\ref{fig:lattice}(c). We will use this lattice to demonstrate the fusion of two $\tt$ excitations (Fig.~\ref{fig:fbt}). 

The next smallest lattice with spherical boundary conditions consists of six edges and four plaquettes and is equivalent to a tetrahedron, as shown in Fig.~\ref{fig:lattice}(d).  We will use this lattice to realize a different implementation of the $\tt$ fusion, which reveals additional details about the  structure of fusion channels, and discuss the difference between the two cases. 

To realize the braiding properties of the DFib model~(see Fig.~\ref{fig:fbt}(b)) we need to create the chiral excitations $\t1$ or $\1t$ without violating the vertex operators $Q_v$. This can be done using the extended tailed lattice. The simplest tailed  lattices are shown in Fig.~\ref{fig:lattice}(e-h). These lattices have additional qubits to account for the addition of a tail to each plaquette. The simplest lattice on which the creation and braiding of two pairs of chiral anyons can be demonstrated is the \textit{tailed $\Theta$} lattice, which consists of nine qubits~(see Fig.~\ref{fig:lattice}(g)).

We will also briefly discuss a simple procedure for revealing the topological twist phase of a single $\t1$ anyon, which is created as part of a pair on the simplest two-plaquette lattice, or a generalized tadpole, consisting of four qubits (Fig.~\ref{fig:fbt}(c)). A simple generalization of this procedure can then be used to verify the relation between braiding and twisting.

\textit{A note on notation:} For the remainder of the paper, we will use two different notations for our states, depending on the context. The first is the qubit basis, in which we simply refer to the states of the qubits ($|0\ra$ or $|1\ra$) that form a particular lattice. The second notation is used when considering the concentric tadpole basis~(see Fig.~\ref{fig:concen}), where  we use a tensor product of kets that represent the sectors of the DFib model corresponding to each tadpole in this basis. 
For example, in the case of the $\Theta$ lattice where the concentric tadpole basis consists of two tadpoles (see Fig.~\ref{fig:concen}(a)), we use the notation $|inn.\ra|out.\ra$ to refer to the states of the \textit{innermost} and \textit{outermost} tadpoles. Likewise for the tetrahedron lattice, where the concentric tadpole basis consists of three tadpoles (See Fig.~\ref{fig:concen}(b) and (c)), we use the notation $|inn.\ra|mid.\ra|out.\ra$, which again refer to specific anyon types of the DFib model represented by each of the innermost, middle and outermost tadpoles in this basis.    

\begin{figure}[t!]
\centering
\includegraphics[width = .49\textwidth]{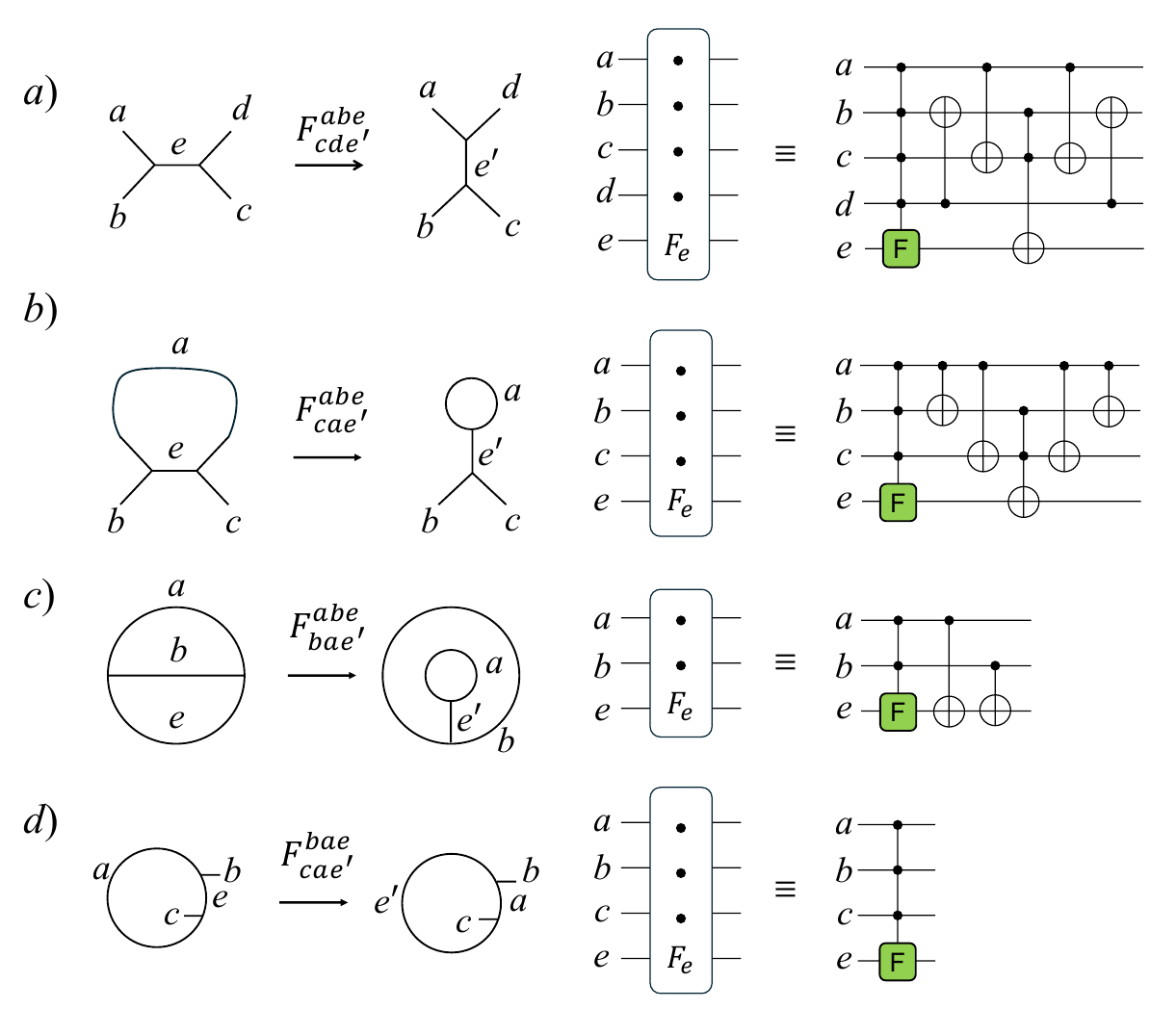}
\caption{
The $F$-move and its circuit implementations. (a) The general form of $F$-move with five inputs. (b) A simplified $F$-move with four inputs. (c) A further simplified $F$-move for three inputs. (d) A special case of $F$-move with four inputs.
}
\label{fig:fmove}
\end{figure}

In what follows, we explain the details of how various excitations can be created, braided, twisted and measured in specific lattices, and 
present the corresponding quantum circuits. The circuit representations of the various $F$-moves used in this section are given in Fig.~\ref{fig:fmove}.

\begin{figure*}[t!]
\centering
\includegraphics[width = .90\textwidth]{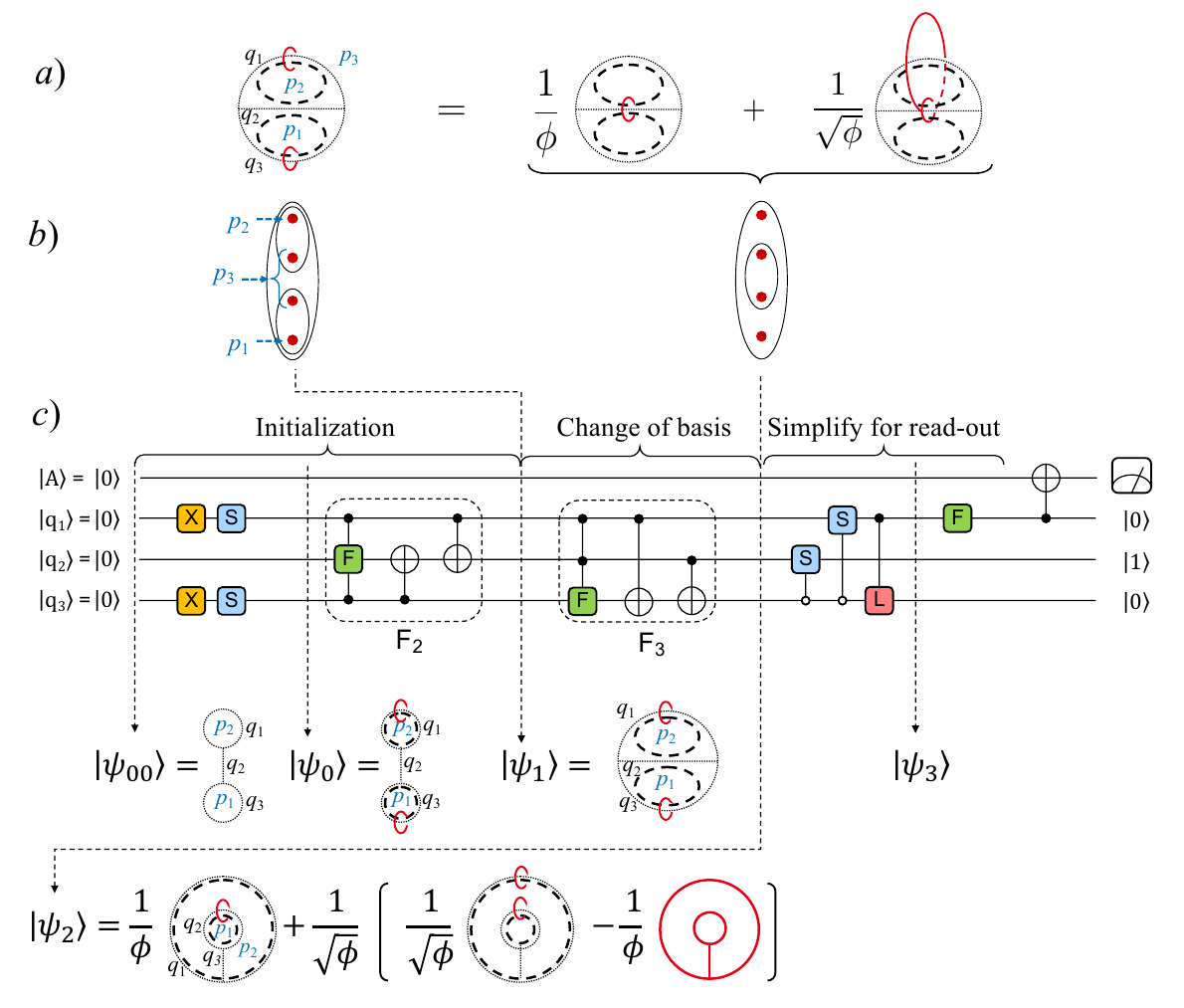}    
\caption{Minimal initialization and measurement circuit for the detection of fusion channels of two $\tt$ excitations. Here we depict qubit state $|q_i\ra = |0\ra$ with dotted line, $|q_i\ra = |1\ra$ with solid red, and the vacuum loop with  dashed line as before. (a)~Graphical representation of creating two pairs of $\tt$ excitations on plaquettes $p_1, p_3$ and $p_2, p_3$, then changing the basis to read the total charge in $p_3$ or equivalently $p_1\cup p_2$. 
(b) A different graphical representation of the same idea where $\tt$ anyons are shown as red dots, created in pairs out vacuum on the left side, then measured in a different basis on the right side. 
(c) The circuit and corresponding state for realizing the initialization and measurement.  
Starting from $|\psi_{00}\ra = |000\ra$, we initialize two $\tt$ excitations on plaquette pairs $p_1,p_3$ and $p_2,p_3$ by first applying the Pauli $X$ (NOT) gate and the $S$ gate to $q_1$ and $q_3$. This effectively inserts two vacuum loops pierced with $\tau$ rings  inside the two tadpoles enclosing $p_1$ and $p_2$, resulting in the state $|\psi_0\ra$. 
We then apply an $F$-move to $q_2$, resulting in $|\psi_1\ra$ in the $\Theta$ lattice basis.
To measure the total charges of $p_1 \cup p_2$, we need to change the basis into concentric tadpoles. For this, we first apply an $F$-move to $q_3$, resulting in $|\psi_2\ra$. Here the outer tadpole (with $q_1$ and $q_3$ as its head and tail, respectively) contains the charge in plaquette $p_3$, or equivalently the total charge of $p_1 \cup p_2$.   
To simplify the detection process, we apply three two-qubit gates to $|\psi_2\ra$, which results in the state $|\psi_3\ra$. These gates are two controlled $S$-gates acting on $q_1$ and $q_2$ when $q_3 = |0\ra$ (note the white control dots) followed by a controlled $L$-gate acting on qubit $q_3$ when $q_1 = |1\ra$. The $S$ and $L$ gates are given in Eq.~\eqref{eq:Smatrix} and~\eqref{eq:Lmatrix}, respectively. At this stage, qubit $|q_1\ra$  contains the information about the fusion channels $\oo$ and $\tt$. 
Here we apply an $F$ gate to $q_1$ as a filter, to map the expected coefficients to the state $|0\ra$ then use it as a control input for a CNOT gate acting on an ancilla qubit $|A\ra = |0\ra$. 
If the state of the ancilla qubit remains unchanged, then we must have had the correct fusion coefficients in $|\psi_3\ra$. This detection process involves only a single measurement.  
An alternative approach to measurement of $|\psi_2\rangle $ is to measure the state of $q_3$.  If $q_3$ is found to be in the $|1\rangle$ state, so should $q_1$ and $q_2$.  If $q_3$ is found to be in the $|0\rangle$ state, then $q_1$ and $q_2$ can be tomographed, and we should find $q_2$ to be in the  $|\tau \tau\rangle_{11}$ state and $q_1$ to be in the $|1 1 \rangle + |\tau \tau\rangle_{11}$ state. 
}
\label{fig:thetafuse_tt}
\end{figure*}

\subsection{Fusion with Three Qubits: The $\Theta$ Lattice}
The simplest setting for detecting the fusion properties of the DFib model is to realize two pairs of $\tt$ excitations on a minimal lattice consisting of three plaquettes on a sphere, or the $\Theta$ lattice. 
The process of initialization and detection of fusion channels is shown in Fig.~\ref{fig:thetafuse_tt}.

\textit{Initialization:}  We start by initializing two pairs of $\tt_{11}$ excitations on plaquette pairs $p_1, p_3$ and $p_2, p_3$. This can be done by starting from a basis in which each of the plaquettes $p_1$ and $p_2$ are reduced to single tadpole, as shown by the trivial state  shown as $|\psi_{00}\ra$ in Fig.~\ref{fig:thetafuse_tt}. Here  qubits $q_1$ and $q_3$ form the heads of the two tadpoles and qubit $q_2$ serves as the common tail.   We then apply single-qubit gates to the two heads, resulting in the state $|\psi_0\ra$, which in qubit basis has the form,
\be
|\psi_0\ra &=& |q_1q_2q_3\ra \\\nn
&=& \left(\frac{\phi|0\ra - |1\ra}{D}\right)|0\ra \left(\frac{\phi|0\ra - |1\ra}{D}\right).
\ee
We then apply an $F$-move to qubit $q_2$  to restore the original $\Theta$ lattice basis, resulting in the state $|\psi_1\ra$.  At this point, we have created two pairs of $\tt$ excitations on plaquette pairs $p_1, p_3$ and $p_2, p_3$ in the $\Theta$ lattice, such that the total charges of plaquettes $p_1\cup p_3$ and $p_2\cup p_3$ correspond to the $\oo$ sector of the DFib model. In other words, we have effectively pulled two  pairs of $\tt_{11}$ excitations out of vacuum.

\textit{Measurement:}  We expect the total charge of plaquettes $p_1 \cup p_2$, (or equivalently the charge of plaquette $p_3$, due to the spherical boundary conditions) to be in a superposition state consistent with the fusion rules of the Fibonacci model. To read this total charge, we need to change the basis into concentric tadpoles (see also Fig.~\ref{fig:concen}(a)) such that, either plaquette $p_1$ or plaquette $p_2$ will be enclosed by a tadpole, which is nested inside an annulus, containing the charge of the other plaquette. This can be done by applying an $F$-move, acting on qubit $q_3$, in $|\psi_2\ra$ as shown in Fig.~\ref{fig:thetafuse_tt}~\footnote{This could also be done with an $F_1$ move resulting in a shape where $p_2$ is enclosed by the central tadpole.}. 

As was discussed in Sec.~\ref{sec:init}, the outer tadpole, which encloses plaquettes $p_1$ and $p_2$, can now be measured to ascertain the total charge of $p_1 \cup p_2$. Note that in this case this outer boundary is also equivalent to the boundary of the single tadpole that encloses plaquette $p_3$ (the remainder of the sphere) and contains the same charge as $p_1 \cup p_2$. Note also that while we use the term \textit{fusion} in referring to combining topological charge, we don't actually fuse excitations in plaquettes $p_1$ and $p_2$. We merely change the basis such that we can measure the superposition state that contains information on fusion amplitudes, should an actual fusion occur.

Using the notation $|\psi_2\ra  = |inn.\ra|out.\ra$, this state can be written as,
\be
\label{eq:theta_fuse_tt}
|\psi_2\ra  &=& \fphi \tt_{11}\oo \\\nn
&+& \fsqphi \Big(\fsqphi \tt_{11}\tt_{11} 
- \fphi\tt_{1\tau}\tt_{\tau1}\Big),
\ee
which represents the anyon types contained by the inner and outer tadpoles, as shown in Fig.~\ref{fig:thetafuse_tt}. We see that the inner tadpole, which contains the charge of plaquette $p_1$, corresponds to a single $\tt$ anyon in all three terms. This is consistent with our initialization, in which plaquettes $p_1$ and $p_2$ each contained a single $\tt$ excitation. 

Focusing on the outer tadpole, which contains the total charge of plaquettes $p_1 \cup p_2$, the first term in $|\psi_2\ra$ indicates trivial total charge with coefficient $1/\phi$, resulting in the state $\oo$, while the second and third terms indicate that the total charge corresponds to the charge of a single $\tt$ anyon with coefficient $1/\sqrt{\phi}$. Thus, the total charge of of $p_1\cup p_2$, signifies the following cross fusion relation,
\be
\tt \times \tt = \fphi \oo + \fsqphi \tt,
\label{eq:fusetheta_exp}
\ee
i.e., only two fusion channels appear. We will return to this point at the end of this section.

Note that while we started by initializing the $\tt_{11}$ component of the $\tt$ excitation, cross fusion results in additional components of the $\tt$ excitation. This can be seen in the third term of $|\psi_2\ra$ (see Fig.~\ref{fig:thetafuse_tt} and Eq.~\eqref{eq:theta_fuse_tt}) where the excitation $\tt_{1\tau1}$ appears in the inner tadpole and $\tt_{\tau1}$ in the outer one. Since the components of $\tt$ contain the same charge, this is consistent with Eq.~\eqref{eq:fusetheta_exp}, where the second term can contain different combinations of the components of $\tt$.  These terms can also be obtained by resolving the loops on the right hand side of the equation shown in Fig.~\ref{fig:thetafuse_tt}(a) followed by the $F$-move.

To simplify the measurement of the fusion channels of $p_1 \cup p_2$ in a quantum circuit, we apply a series of two-qubit gates to $|\psi_2\ra$, which effectively maps it to a product state. These two-qubit gates are two controlled-$S$ gates acting on qubits $q_2$ and $q_3$ and a controlled-$L$ gate, acting on qubit $q_3$ where,
\be
\label{eq:Lmatrix}
L = \begin{pmatrix}
\frac{1}{\sqrt{\phi}} & -\frac{1}{\phi}   \\
-\frac{1}{\phi}  & -\frac{1}{\sqrt{\phi}} \\
\end{pmatrix},
\ee
resulting in the state
\be
|\psi_3\ra &=& |q_1q_2q_3\ra \\\nn
&=&  \big(\fphi|0\ra + \fsqphi|1\ra \big)|10\ra.
\ee
Thus, the information on the fusion channels of  $p_1 \cup p_2$ is effectively transferred to qubit $q_1$ in this state. 

To detect the fusion coefficients, we can perform statistical measurements on $q_1$ at this stage and observe the corresponding probability distribution. Alternatively, we can apply an $F$ gate (Eq.~\eqref{eq:Fmatrix}) to $q_1$, then use the result as the control for a CNOT gate acting on a target ancilla qubit. This $F$ gate acts as a ``filter,'' so that if $q_1$ is in the correct superposition, the $F$ gate maps it to $|0\ra$, which does not affect the ancilla qubit in the subsequent CNOT. Thus, with this protocol we can verify the coefficient of fusion by carrying out a single measurement of the ancilla qubit. 

Alternatively, we can determine the state of  $|\psi_2\rangle$ by first measuring the state of $q_3$.  If $q_3$ is found to be in the $|1\rangle$ state, so should $q_1$ and $q_2$ (the last term in the expression for $|\psi_2\rangle$ in the Fig ~\ref{fig:thetafuse_tt}.c).  If $q_3$ is found to be in the $|0\rangle$ state, then $q_1$ and $q_2$ can be tomographed, and we should find $q_2$ to be in the  $|\tau \tau\rangle_{11}$ state and $q_1$ to be in the $|1 1 \rangle + |\tau \tau\rangle_{11}$ state, which directly reflects the fusion in Eq.~\ref{eq:fusetheta_exp}. In cases below we will always favor read-out schemes that minimize the number of measurements, although other read-out schemes are certainly possible.

To conclude this section, we return to the observation in ~Eq.~\eqref{eq:fusetheta_exp}, namely the fact that in the $\Theta$ lattice, the result of cross fusing two $\tt$ excitations is two fusion channels, namely $\oo$ and $\tt$. This might look unexpected at first sight, since the two sets of chiral excitations in the DFib model (the right-handed $\t1$ and the left-handed $\1t$) must fuse independently from each other, thus in principle we expect the combination of two $\tt$ excitations to result in four fusion channels: $\oo$, $\t1$, $\1t$ and $\tt$. However, recall that $\t1$ and $\1t$ are created by passing a $\tau$ string over or under the lattice, respectively. On the $\Theta$ lattice, two excitations must share the same plaquette and for a single plaquette there is no way to distinguish between ``over'' and ``under.'' Thus only the achiral excitations can exist on this plaquette. Consequently,  the combined charge of $p_1\cup p_2$ lead to only two  fusion  channels: $\oo$ and $\tt$.

\begin{figure*}[t!]
\centering
\includegraphics[width = \textwidth]{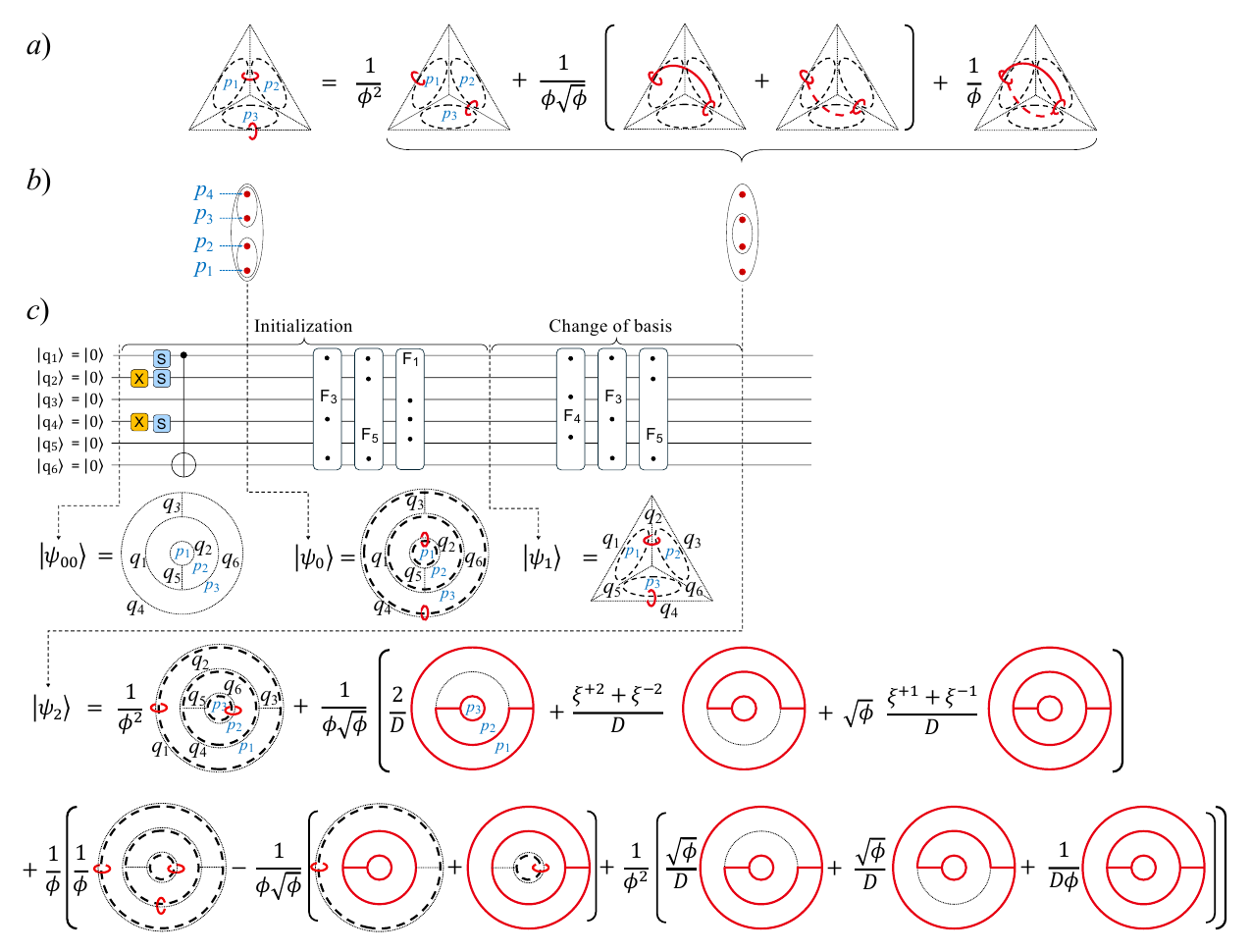}
\caption{(a) Graphical representation of the fusion channels of two $\tt$ excitations in the smallest lattice without tails in which each plaquette contains a single $\tt$ excitation. Here we depict qubit state $|q_i\ra = |0\ra$ with dotted line, $|q_i\ra = |1\ra$ with solid red, and the vacuum loop with  dashed line as before. The left side of the equation represents two pairs of $\tt$ excitations on plaquettes $p_1, p_2$ and $p_3, p_4$, each pair with trivial total charge, corresponding to $\oo$. The right hand side shows this state in a different basis, where the total charges of $p_2 \cup p_3$ are measured. 
(b) A different graphical representation of the same idea where $\tt$ anyons are shown as red dots, created in pairs out vacuum on the left side, then measured in a different basis on the right side. 
(c) The circuit for initialization and change of basis to read-out. We start by a concentric tadpole basis in which plaquette $p_1$ is enclosed by the innermost tadpole, followed by $p_2$ represented by an annulus surrounding $p_1$ and $p_3$ forming the next annulus enclosed by the outer tadpole. The trivial state in this basis is shown as $|\psi_{00}\ra$. We initialize this state by carrying out the NOT, $S$ (Eq.~\eqref{eq:Smatrix}) and CNOT gates on qubits forming the heads of the concentric tadpoles, resulting in $|\psi_0\ra$. 
We then apply three $F$-moves to qubits $q_3$, $q_5$ and $q_1$ to return  to the tetrahedron lattice basis, resulting in $|\psi_1\ra$. At this point, we have created two pairs of $\tt$ excitations on plaquettes $p_1, p_2$ and $p_2, p_3$.
To measure the total charge of $p_2\cup p_3$, we carry out another change of basis, by applying $F$-moves to qubits $q_4$, $q_3$, and $q_5$, resulting in $|\psi_2\ra$, which is written in a different concentric tadpole basis. In this basis, the total charge of plaquettes $p_2\cup p_3$ is enclosed by the middle generalized tadpole, where qubits $q_2$ and $q_4$ form the head, and qubits $q_3$ and $q_5$ form the two tails. Thus, these four qubits collectively contain the information on the charge content and fusion channels of $p_2 \cup p_3$. 
The middle tadpole (representing the charge of $p_2\cup p_3$) in the first term of $|\psi_2\ra$, with coefficient $1/\phi^2$, indicates a total charge of $\oo$ and corresponds to the first term on the right hand side of (a). 
The next three terms of $|\psi_2\ra$, with overall coefficient $1/\sqrt{\phi}$, represent the sum of chiral excitations $\t1$ and $\1t$, corresponding to the middle two terms on the right hand side of (a). 
Finally, the terms in the second line of $|\psi_2\ra$, with overall coefficient $1/\phi$, represent the charge of $\tt$, corresponding to the last term in (a). Note that the charges of the innermost tadpole (representing $p_3$) and the outermost tadpole (representing $p_1 \cup p_2 \cup p_3$) remain at $\tt$. 
}
\label{fig:fusetetra}
\end{figure*}

To explore this idea further, we create two pairs of $\tt_{11}$ excitations on the four plaquettes of a  tetrahedral lattice, consisting of six qubits, and measure the cross-fusion channels of opposite plaquettes. We expect to see four fusion channels in this case, including the chiral excitations. We will discuss this in the next section. 
\subsection{Fusion with Six Qubits: Tetrahedron Lattice}
The process of initialization and measurement of the fusion channels of two pairs of excitations on the four plaquettes of a tetrahedral lattice is shown in Fig.~\ref{fig:fusetetra}. 

\textit{Initialization:}  We start by initializing two pairs of $\tt$ excitations out of vacuum on pairs of plaquettes $p_1, p_2$ and $p_3, p_4$ such that the total charge of each pair corresponds to $\oo$. This can be done by starting from a basis in which plaquettes $p_1$ and $p_2$ form concentric tadpoles, enclosed by a third tadpole containing $p_3$. In Fig.~\ref{fig:fusetetra}, one such starting configuration is depicted as the basis for the state $|\psi_{00}\ra$. We then create pairs of $\tt_{11}$ excitations on plaquettes $p_1, p_2$ and $p_3, p_4$ by initializing the corresponding generalized tadpoles. 

The innermost tadpole, which represents plaquette $p_1$, consists of qubits $q_2$ and $q_5$, forming its head and tail, respectively. This tadpole is initialized as, 
\be
|q_2 q_5\ra = \frac{\phi|0\ra - |1\ra}{D}|0\ra.
\ee
The middle (generalized) tadpole, which contains the information about total charges of $p_1 \cup p_2$, consists of qubits $q_1$ and $q_6$,  together forming its head, and $q_5$ and $q_3$, which form its two tails. This tadpole is initialized as,
\be
|q_1 q_6 q_5 q_3\ra = \frac{|00\ra + \phi |11\ra}{D}|00\ra
\ee
Finally, the outermost tadpole, containing the charge of $p_1 \cup p_2 \cup p_3$, consists of qubit $q_4$ as its head and qubit $q_3$ as its inward tail. These qubits are initialized as,
\be
|q_4 q_3\ra = \frac{\phi|0\ra - |1\ra}{D}|0\ra.
\ee
The resulting state in the qubit basis is the following,
\be
|\psi_0\ra &=& |q_2 q_4\ra |q_1 q_6\ra |q_5 q_3\ra \\\nn
&=& 
\left(\frac{\phi|0\ra - |1\ra}{D} \right)^{\otimes2}\left(\frac{\phi|00\ra +\phi |11\ra}{D}\right)|00\ra.
\ee
We can also re-write the state $|\psi_0\ra$ in the concentric tadpole basis, where each ket represents the topological charge of the corresponding generalized tadpole,
\be 
|\psi_0\ra &= |inn.\ra |mid.\ra |out.\ra 
\ee
where,
\be
\begin{aligned}
|inn.\ra &= \tt_{11}\\\nn
|mid.\ra &= \oo\\\nn
|out.\ra &= \tt_{11}.
\end{aligned}
\ee

Here the first equality means that the plaquette represented by the innermost tadpole ($p_1$) and the plaquette represented by the annulus surrounding the innermost tadpole ($p_2$) each have a charge of $\tt_{11}$. The second equality indicates that the total charge of plaquettes $p_1 \cup p_2$, represented by the middle tadpole, is $\oo$; in other words, the pair of excitations on $p_1$ and $p_2$ are effectively pulled out of vacuum. Finally, the third equality implies that the plaquette represented by the annulus surrounding the middle tadpole ($p_3$) and the outside plaquette representing the remainder of the sphere ($p_4$) each also have a charge of $\tt_{11}$. Due to spherical boundary conditions, the total charge of these two plaquettes is also $\oo$, thus the excitations on these plaquettes are also pulled out of vacuum. 

To return to the original lattice basis (tetrahedron) we apply three $F$-moves to qubits $q_3$, $q_5$ and $q_1$, as shown in Fig.~\ref{fig:fusetetra}(b), resulting in the state $|\psi_1\ra$. The circuit definitions of the $F$-moves can be found in Fig.~\ref{fig:fmove}. At this point, we have created two pairs of $\tt$ excitations on plaquettes $p_1, p_2$ and $p_3, p_4$ such that each pair fuse to vacuum.

\begin{figure*}[t!]
\centering
\includegraphics[width = .95\textwidth]{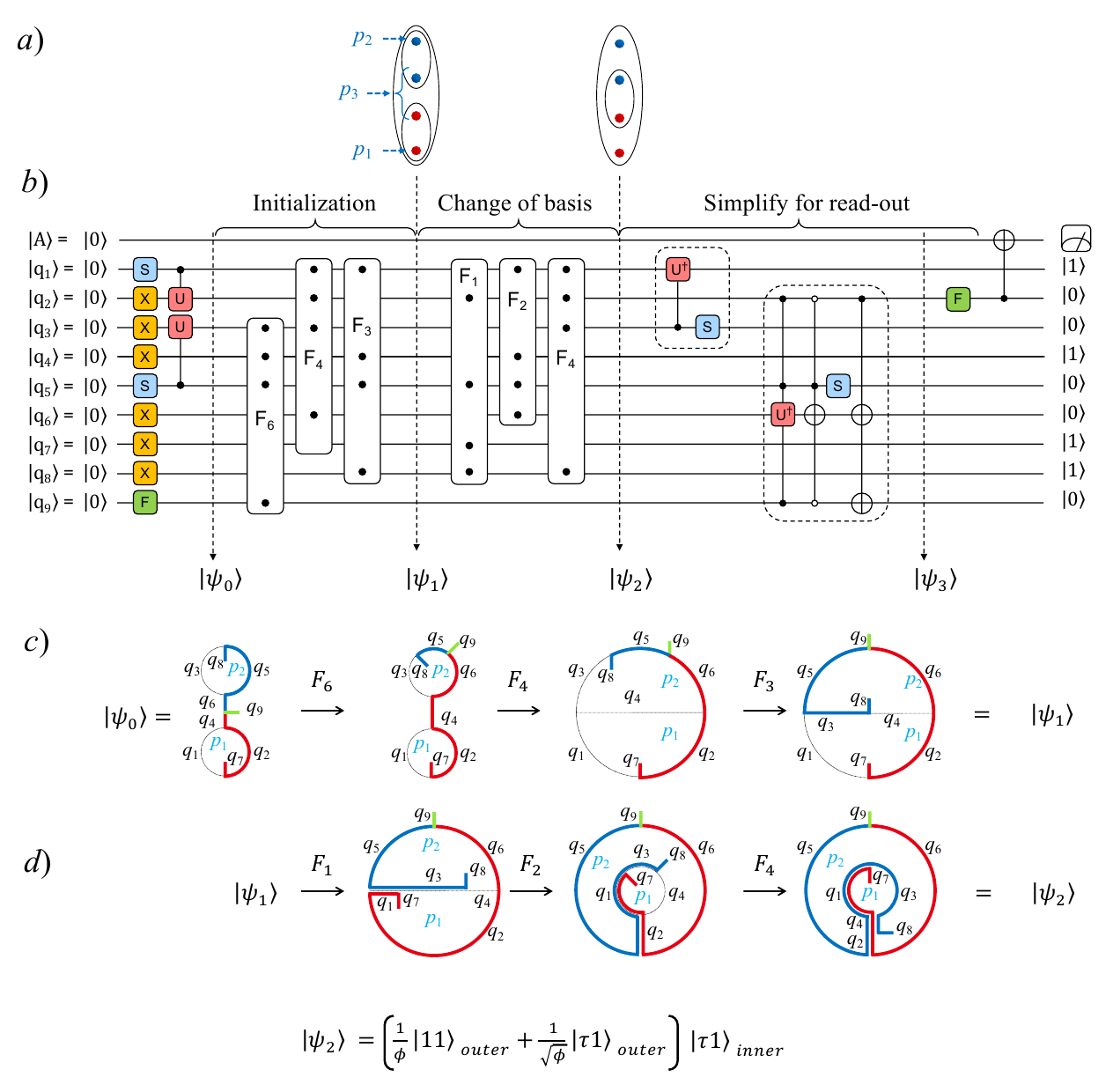}
\caption{(a) Schematic depiction of creating two pairs of anyons, then reading the cross-fusion charges in the read-out basis. 
(b) Quantum circuit for the creation and measurement of the two fusion channels of two $\t1$ excitations. 
Starting from a basis where plaquettes $p_1$ and $p_2$ form two tadpoles, we create two pairs of $\t1$ excitations $p_1, p_3$ and $p_2, p_3$, by initializing the tadpoles using the circuits given in Fig.~\ref{fig:init}. Note that we initialize the tail of plaquette $p_3$ with an $F$-move. An alternative procedure is described in the text. We then apply three $F$-moves to return to the tailed $\Theta$ lattice basis, resulting in the state $|\psi_1\ra$. 
In preparation for read-out, we then apply three additional $F$-moves which maps the state to the concentric tadpole basis, resulting in $|\psi_2\ra$. At this point we have effectively pulled two pairs of $\t1$ excitations out of vacuum such that plaquette $p_3$ contains two anyons. 
In preparation for read-out, we apply three additional $F$-moves to change the basis into concentric tadpoles such that the outer tadpole contains the total charge of $p_1\cup p_2$, or equivalently $p_3$, resulting in the state $|\psi_2\ra$. At this point the outer tadpole, consisting of $q_5$ and $q_6$ as its head and $q_2$ and $q_9$ as its two tails, contains all the information on the fusion channels of the combined charges on $p_1\cup p_2$.
To simplify the read-out, we apply additional gates from Fig.~\ref{fig:init}. Here the first block of gates acts on the inner tadpole and the second block on the outer tadpole. These gates  map $|\psi_2\ra$ to a non-topological un-entangled state, $|\psi_3\ra$ (see the text for the details). Qubit $q_2$ in this state now contains all the information on the fusion channels. We can carry out a statistical measurement on $q_2$ at this stage to determine the fusion coefficients, or alternatively apply an $F$ gate as a filter to verify the fusion coefficients. 
(c) The change of basis from the initialization basis to the $\Theta$ lattice basis, where the effects of the $F$-moves are shown on the lattice. The blue and red lines keep track of the locations of the anyons. The green line represents the state of plaquette $p_3$’s tail. The solid red and blue lines are only guides to the eye. 
(d) The change of basis from the $\Theta$ lattice basis to the concentric tadpole basis. Here again the effects of the $F$-moves on the lattice are shown, and the blue and red lines keep track of the locations of the $\t1$ anyons. 
}
\label{fig:fuse_t1}
\end{figure*}

\textit{Measurement:} To measure the fusion coefficients resulting from cross-fusing excitations from opposite pairs, e.g., $p_2 \cup p_3$ and $p_1 \cup p_4$, we need to switch to a concentric tadpole basis in which two plaquettes from opposite pairs are enclosed by the middle tadpole. One such basis is shown in Fig.~\ref{fig:concen}(c), in which the middle tadpole encloses the charge of $p_2\cup p_3$. In this basis, the innermost tadpole contains the charge of plaquette $p_1$, and the outermost tadpole corresponds to the total charge of plaquettes $p_1 \cup p_2 \cup p_3$ or, equivalently, plaquette $p_4$. This basis change can be achieved by applying three additional $F$-moves acting on qubits $q_4$, $q_3$ and $q_5$,  resulting in the state $|\psi_2\ra$ in Fig.~\ref{fig:fusetetra}. In the concentric tadpole notation, this state has the form,
\begin{widetext}
\be
\begin{aligned}
|\psi_2\ra &= \frac{1}{\phi^2} \tt_{11}\oo\tt_{11} \\
&+ \frac{1}{\phi\sqrt{\phi}} \tt_{1\tau} \Big(\t1 + \1t\Big) \tt_{\tau1} \\
&+\frac{1}{\phi} \Bigg[
\frac{1}{\phi}\tt_{11}\tt_{11}\tt_{11} 
-\frac{1}{\phi\sqrt{\phi}} \Big( \tt_{\tau1}\tt_{1\tau}\tt_{11} + \tt_{11}\tt_{1\tau}\tt_{\tau1}\Big) 
+ \frac{1}{\phi^2} \tt_{\tau1}\tt_{\tau\tau}\tt_{1\tau}\Bigg].
\end{aligned}
\label{eq:tetra_full}
\ee
\end{widetext}

As was noted above, the information on the fusion channels of the total charge of $p_2\cup p_3$ is now represented by the state of the middle tadpole in $|\psi_2\ra$. To identify the components of the fusion channels, we can use Eq.~\eqref{11}-\eqref{tttt} to re-write $|mid.\ra = \tt \times \tt $ as follows,
\be
\label{eq:tetrafuse}
|mid.\ra 
&=& 
\frac{1}{\phi^2} \oo \\\nn
&+& 
\frac{1}{\phi\sqrt{\phi}}(\t1  + \1t)  \\\nn 
&+& \fphi \tt,
\ee
where,
\be
\label{fig:fusetetra_tttt}
\tt &=& \fphi \tt_{11} \\ \nn
&-& \frac{1}{\phi\sqrt{\phi}}\Big( \tt_{1\tau} + \tt_{\tau 1}\Big)\\ \nn
&+&\frac{1}{\phi^2}\tt_{\tau\tau}.
\ee
This is consistent with our expectation that in general the two chiral sectors of the $\tt$ excitation fuse independently, resulting in four fusion channels, as given in Eq.~\eqref{eq:tetrafuse}.

The charges of individual plaquettes $p_3$ (represented by the innermost tadpole $|inn.\ra$) and $p_4$ (represented by the outermost tadpole $|out.\ra$) remain at $\tt$. However, as we see in Eq.~\eqref{eq:tetra_full}, for these states different components of the $\tt$ sector appear in different terms. 

Similar to the previous case, it is possible to design a quantum circuit to simplify the state $|\psi_1\ra$ such that the expected configuration will be mapped to an un-entangled state which can then be measured for verification.

\subsection{Fusion and Braiding with Nine Qubits: Tailed $\Theta$ Lattice}
\label{sec:fuse-braid-theta}
To detect non-Abelian braiding phases in the DFib model, we need to create the chiral excitations $\t1$ or $\1t$. To realize these excitations without violating the vertex operators $Q_v$, we employ the extended tailed lattice~\cite{feng15, hu18, schotte20}. The simplest nontrivial tailed lattice for this purpose would be a tailed $\Theta$ lattice, consisting of nine qubits, as shown in Fig.~\ref{fig:lattice}(g). In what follows we describe circuits to create these excitations, change the basis to measure fusion channels, then braid the excitations and finally detect the braiding phases.

\begin{figure}[t]
\centering
\includegraphics[width = .49\textwidth]{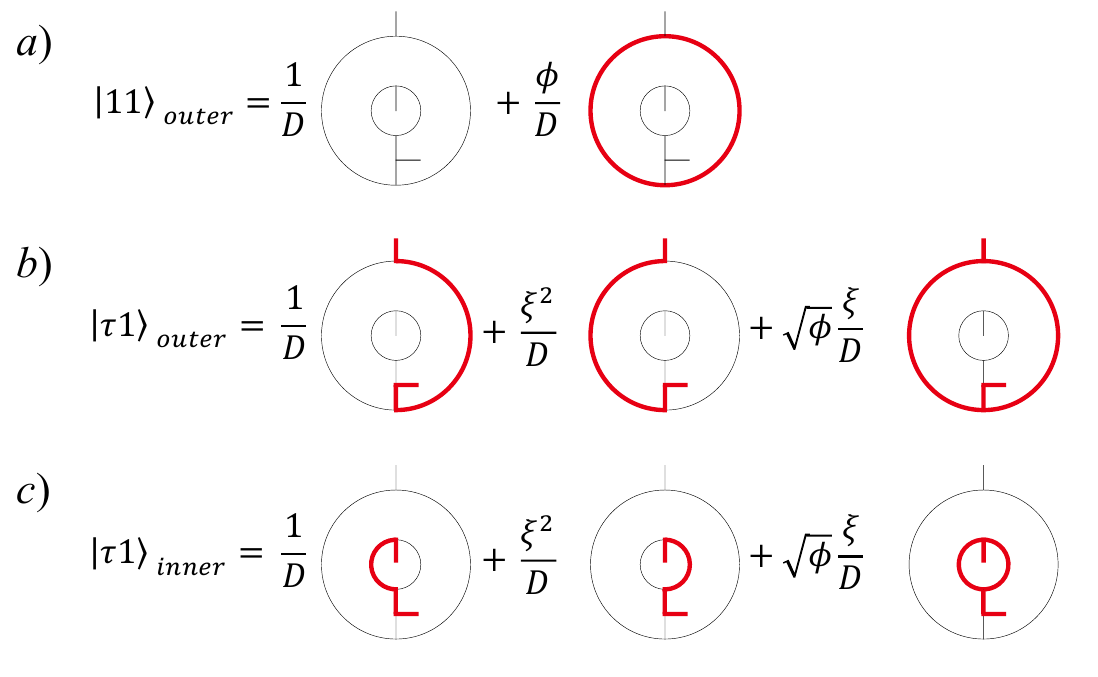}
\caption{The explicit forms of the expected states of the inner and outer tadpoles for a tailed $\Theta$ lattice. (a) The configuration corresponding to the outer tadpole carrying the charge of $\oo$. (b) Outer tadpole carrying the charge of $\t1$. (c) Inner tadpole carrying the charge of $\t1$.}
\label{fig:t1_tadpole_basis}
\end{figure}

\begin{figure*}[t!]
\centering
\includegraphics[width = .8\textwidth]{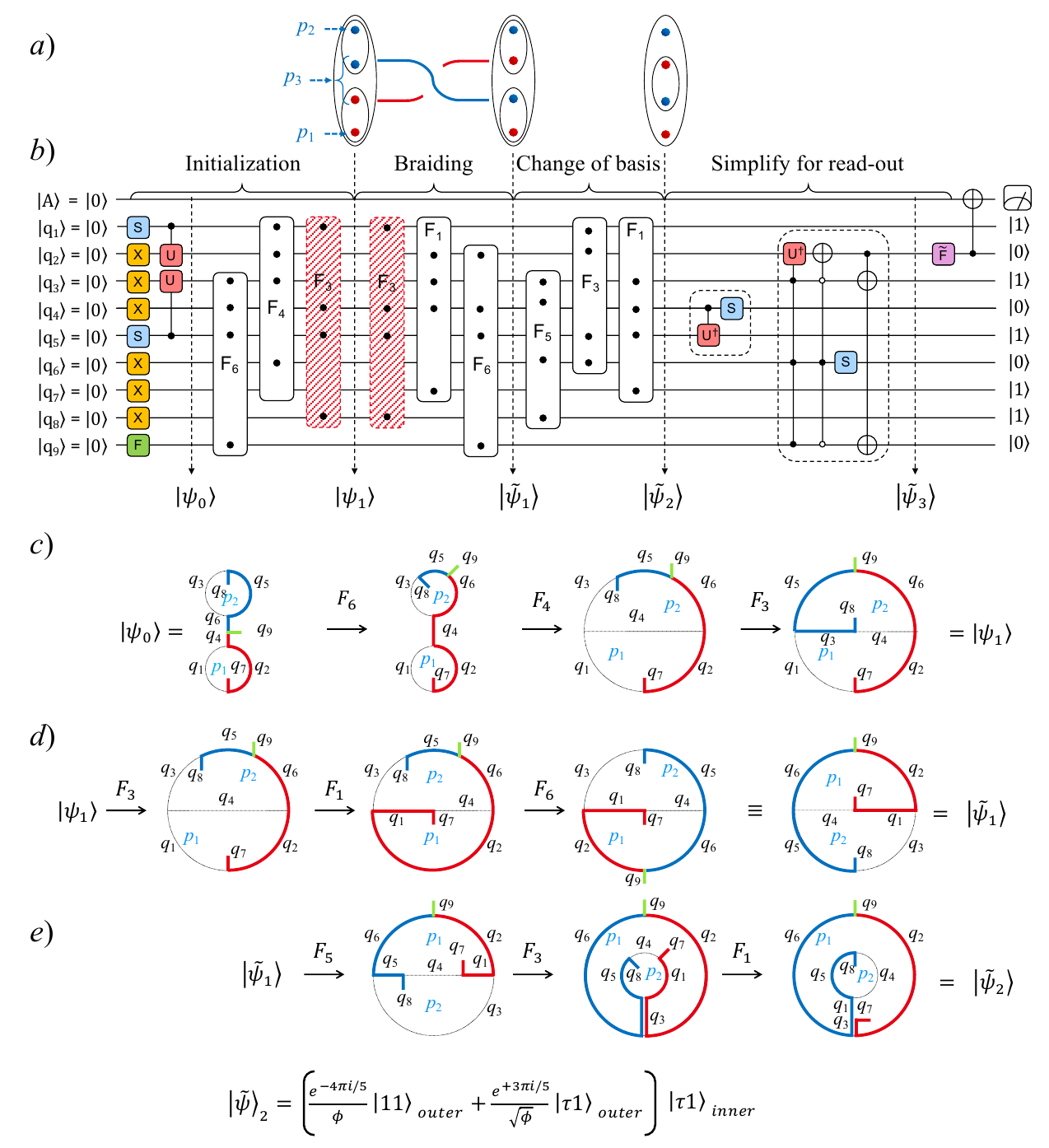}
\caption{
(a) Schematic depiction of creation, braiding and read-out of two $\t1$ excitations. 
(b) The full quantum circuit from initialization to final measurement. The $F_3$-moves shown in red, which occur at the end of the initialization and the beginning of braiding steps, are shown for completeness of each step, but they cancel each other and in practice can both be removed from the circuit. Here $X$ is the NOT gate, and the matrix representations of the $S$, $F$, $U$ and ${\tilde F}$ gates are given in the text. 
We start by creating two pairs of $\t1$ excitations on plaquettes $p_1, p_3$ and $p_2, p_3$. We then apply three $F$-moves to restore the lattice basis. This initialization up to $|\psi_1\rangle$ is identical to that of Fig.~\ref{fig:fuse_t1}. 
We then carry out a single braid, by exchanging plaquettes $p_1$ and $p_2$ in a counterclockwise manner with the aid of three $F$-moves, acting on qubits $q_3$, $q_1$ and $q_6$, resulting in the state $|\tilde\psi_1\ra$.
To facilitate read-out, we apply three $F$-moves to change the basis into concentric tadpoles such that the outer tadpole contains the total charge of $p_1\cup p_2$, or equivalently $p_3$, resulting in the state $|\tilde\psi_2\ra$. At this point the outer tadpole, consisting of $q_2$ and $q_6$ as its head and $q_3$ and $q_9$ as its two tails, contains all the information on the fusion channels and braiding phases of the combined charges on $p_1\cup p_2$.
To facilitate read-out, we simplify $|\tilde{\psi}_2 \ra$ bu applying two sets of gates. Here the first block of gates acts on the inner tadpole and the second block on the outer tadpole.
These gates reduce the state to a product state such that the braiding phases will be stored in the superposition state of a single qubit, in this case $q_2$. This state then passes through the filter ${\tilde F}$ (Eq.~\eqref{eq:F_tilde}), mapping it to $|0\ra$, which subsequently acts as a control qubit for a CNOT acting on the ancilla qubit $|A\ra = |0\ra$. An unchanged ancilla indicates the expected braiding phases. 
(c) The initialization of two pairs of $\t1$ excitations, identical to the case of fusion Fig.~\ref{fig:fuse_t1}.
(d) Effecting one counterclockwise exchange of the excitations on plaquettes $p_1$ and $p_2$ in the $\Theta$ lattice basis, using three $F$-moves.  Note  that we have rotated the  last lattice in this diagram by $\pi$ to restore the original orientation of the tails. Here again the effects of the $F$-moves on the lattice are shown and the blue and red lines keep track of the locations of the $\t1$ anyons. The braiding can be understood by comparing the diagram representing $|\psi_1\ra$ in (c) with the diagram representing $|\tilde\psi_1\ra$ and notice that plaquettes $p_1$ and $p_2$, and their respective tail qubits $q_7$ and $q_8$ have switched places compared.
(e) The details of changing the basis into concentric tadpoles to prepare for measurement. The braiding phases are accumulated in the state of the outer tadpole, which contains the total charges of $p_1 \cup p_2$. The resulting state is written in the concentric tadpole basis, $|\tilde{\psi}_2 \ra = |inn.\ra|out.\ra$. 
}
\label{fig:braid_t1}
\end{figure*}
%
\subsubsection{Fusion}
Here we focus on creating two pairs of $\t1$ excitations on three plaquettes, aiming at detecting the following fusion relation,
\be
\t1 \times \t1 = \fphi \oo + \fsqphi \t1,
\ee
which results from fusing excitations from opposite pairs, created out of vacuum. The general procedure resembles the case of creating $\tt$ excitations in the $\Theta$ lattice described above, but the initialization and measurement processes involve details that are specific to the chiral excitations, which we describe bellow. The full circuit is depicted in Fig.~\ref{fig:fuse_t1}.

\textit{Initialization:} We initialize the tailed $\Theta$ lattice in a basis in which plaquettes $p_1$ and $p_2$ are enclosed by separate (generalized) tadpoles, resulting in $|\psi_0\ra$, as shown in Fig.~\ref{fig:fuse_t1}(c). The general form of initialization circuit for  $\t1$ anyons is given in Fig.~\ref{fig:init}. Here we apply these circuits independently to the two tadpoles, but initialize the tail of plaquette $p_3$, with an $F$-gate. This can be understood by first imagining two tails for plaquette $p_3$ by using two extra qubits $q_{10}$ and $q_{11}$, then initialize each tadpole independently according to~Fig.~\ref{fig:init}. We then combine these tails by applying an $F$-move to $q_9$ and through away the extra qubits  $q_{10}$ and $q_{11}$, leaving us with $|q_9\ra=\fphi|0\ra + \fsqphi|1\ra$. Since everything about this process is deterministic, we choose to do away with the extra qubits and directly initialize the tail qubit $q_9$ with an $F$ gate. 

We then apply three $F$-moves to $|\psi_0\ra$ to return the state to the $\Theta$ lattice basis, resulting in $|\psi_1\ra$. This process is depicted in Fig.~\ref{fig:fuse_t1}(c) where the effects of the $F$-moves on the lattice are shown. Here the colored solid lines are merely guides to the eye, to keep track of the locations of pairs of excitations. At this point we have created two pairs of $\t1$ excitations on plaquettes $p_1, p_3$ and $p_2, p_3$, such that each of these pairs would fuse to the vacuum state $\oo$.

\textit{Measurement:} To measure the  amplitudes of the superposition state $|\psi_1\ra$, resulting from cross-fusing the charges of plaquettes $p_1\cup p_2$, we change into a concentric tadpole basis where plaquettes $p_1$ and  $p_2$ are enclosed by concentric tadpoles. This can be done with the aid of three $F$-moves, resulting in the state $|\psi_2\ra$ as shown in Fig.~\ref{fig:fuse_t1}(d). 

In this state, the outer tadpole, consisting of qubits  $q_5$ and $q_6$, which form the head, and qubits $q_2$ and $q_9$, forming its tails, contains the information on the total charge content of plaquettes $p_1 \cup p_2$, or equivalently, $p_3$. Likewise, qubits $q_1$, $q_3$, $q_4$ and $q_7$, which form the components of the inner tadpole, contain information on the charge content of plaquette $p_1$. Written in the concentric tadpole basis, $|\psi_2\ra = |inn.\ra|out.\ra$, we have, 
\be 
|\psi_2\ra = 
\t1\Big(\frac{1}{\phi}\oo +
\frac{1}{\sqrt{\phi}}\1t \Big).
\ee
The components of the $\oo$ and $\t1$ sectors for the inner and outer tadpoles are shown in Fig.~\ref{fig:t1_tadpole_basis}. In terms of these components, the inner tadpole consists of three terms (Fig.~\ref{fig:t1_tadpole_basis}(c)) and the outer tadpole consists of two terms for the $\oo$ sector (Fig.~\ref{fig:t1_tadpole_basis}(a)) and three terms for the $\t1$ sector (Fig.~\ref{fig:t1_tadpole_basis}(b)), leading to fifteen terms for $|\psi_2\ra$.

To detect the fusion amplitudes, we  first simplify $|\psi_2\ra$ using the  circuits given in Fig.~\ref{fig:init} applied independently to the inner and outer tadpoles. Switching to the qubit basis, these circuits first map the two tadpoles to the following:
\be
|q_4 q_8 q_7 q_1 q_3\ra &=& |11110\ra\\\nn
|q_9 q_2 q_6 q_5\ra &=& \fphi|0000\ra + \fsqphi |1110\ra.
\ee
To further simplify, we apply CNOT gates to $q_2$, $q_6$ and $q_9$ to un-entangle them, resulting in 
\be
|\psi_3\ra &=& |q_1 q_2 q_3 q_4 q_5 q_6 q_7 q_8 q_9\ra \\\nn &=&  |1\ra\big(\fphi|0\ra + \fsqphi |1\ra\big)|0100110\ra.
\ee
Thus, the information on the fusion channels of the combined plaquettes $p_1 \cup p_2$ is now stored in the state of a single qubit, namely $q_2$~\footnote{By choosing different combinations of CNOTs we could choose any of $q_2$, $q_6$ or $q_9$ as our flag qubit.}. 

To detect these amplitudes, as before, we either carry out statistical measurements of $q_2$ and measure the probability distribution of each fusion channel, or apply an $F$ gate to $q_2$ as a ``filter,'' mapping it to $|0\ra$, then use the result to control a CNOT acting on ancilla qubit $A$. If the fusion channels are as expected, the ancilla qubit should not change.

\subsubsection{Braiding}
Staying within the tailed $\Theta$ lattice, here we discuss how to create and braid two $\t1$ excitations and measure the resulting phases. This process is shown in Fig.~\ref{fig:braid_t1}.

\textit{Initialization:} We repeat the same initialization as the one we used in the previous section and create two pairs of $\t1$ excitations, on plaquettes $p_1, p_3$ and $p_2, p_3$, resulting in the state  $|\psi_1\ra$ in the $\Theta$ lattice basis.

\textit{Braiding:} To braid the charge contents of plaquettes $p_1$ and $p_2$, we simply exchange the corresponding plaquettes in a counterclockwise manner using a series of $F$-moves~\cite{kkr, schotte20} as shown in Fig.~\ref{fig:braid_t1}(d), resulting in the state $|\tilde{\psi}_1\ra$. Note that, while we end up in the same lattice basis as the starting point, braiding effectively changes the arrangement of qubits on the lattice. We need to keep this in mind when preparing the state for read-out. 

\textit{Measurement:} To measure the braid phases, we again need to change the basis into one where plaquettes $p_1$ and $p_2$ form concentric tadpoles, so that the total charges of combined plaquettes $p_1 \cup p_2$ will be enclosed by the outer tadpole. This process is the same as the one in the fusion case but with different qubits, as shown in Fig.~\ref{fig:braid_t1}(e). Using the concentric tadpole basis,   $|\tilde\psi_2\ra = |inn.\ra|out.\ra$, we get,
\be
|\tilde{\psi}_2\ra = \t1 
\big( \frac{e^{-4\pi i/5}}{\phi} \oo + \frac{e^{+3\pi i/5}}{\sqrt{\phi}} \t1
\big).
\ee 
Thus the two states of the outer tadpole have acquired the phases corresponding to a counterclockwise exchange of two $\tau1$ anyons.

To verify these phases with a quantum circuit, we begin by simplifying $|\tilde\psi_2\ra$, using the detection circuits in Fig.~\ref{fig:init}. This simplification effectively maps the states of the two tadpoles to the following, 
\be
|q_1 q_7 q_8 q_5 q_4 \ra &=& |11110\ra \\\nn
| q_3 q_9 q_2 q_6\ra &=&  \frac{e^{-4\pi i/5}}{\phi}|0000\ra +\frac{e^{+3\pi i/5}}{\sqrt{\phi}}|1110\ra.
\ee
To further simplify this state, we un-entangle qubits $q_2$, $q_3$ and $q_9$ by applying two CNOT gates, resulting in 
\be
|\tilde{\psi}_3\ra &=& |q_1 q_2 q_3 q_4 q_5 q_6 q_7 q_8 q_9\ra \\\nn 
&=& 
|1\ra
\big(\frac{e^{-4\pi i/5}}{\phi}|0\ra + \frac{e^{+3\pi i/5}}{\sqrt{\phi}} |1\ra\big) |1010110\ra.
\ee
Thus, all the information on braiding phases of the two fusion channels is now stored in $q_2$. For the final detection, similar to the previous cases, we design another filter, which has the form,
\be
\label{eq:F_tilde}
\tilde{F} = \begin{pmatrix}
\frac{\xi^2}{\phi} &  \frac{\xi}{\sqrt{\phi}} \\
\frac{\xi^{-1}}{\sqrt{\phi}}  & -\frac{\xi^{-2}}{\phi}. \\
\end{pmatrix}
\ee

We believe this is the simplest setting in which to demonstrate the braiding properties of the chiral excitations of the DFib model. One may also perform these exercises on a tailed tetrahedron (Fig.~\ref{fig:lattice}(h)) with fourteen qubits, such that each of the four plaquettes contains a single $\t1$ excitation. The results are identical to the three plaquette lattice for both fusion and braiding. 
Since this article is focused on minimal models, we do not provide that circuit here. 

\begin{figure}[t!]
\centering
\includegraphics[width = .49\textwidth]{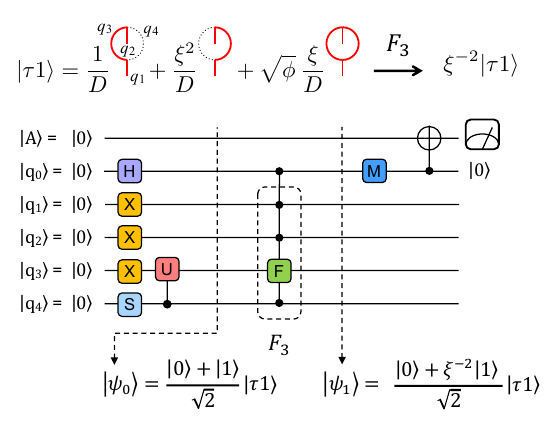}
\caption{Counterclockwise twisting and measurement of a single $\t1$ excitation.}
\label{fig:twist}
\end{figure}

\subsection{Twisting with Five Qubits: Tailed Tadpole}
\label{sec:twist}
A chiral excitation can be twisted within a plaquette and the resulting phase can be detected using additional tricks~(see, e.g., \cite{steve23}). The simplest lattice for this purpose consists of only four qubits, forming a single generalized tadpole, which can be thought as representing two plaquettes on a sphere, one shown and the other being the outside region representing the remainder of the sphere (see Fig.~\ref{fig:lattice}(f)). The circuit for the initialization, twisting and detection of the phase is shown in Fig.~\ref{fig:twist} (with two additional qubits to simplify readout).

We start by initializing  the $\t1$ state on the four qubits forming the generalized tadpole, using the initialization circuits in Fig.~\ref{fig:init}. We then apply a special case of an $F$-move, which effectively exchanges the two tails. This $F$-move is defined in Fig.~\ref{fig:fmove}. Written in the original basis, this is equivalent to twisting the $\t1$ excitation in the front plaquette in a counterclockwise manner, resulting in,
\be
\hat{\theta} \t1 = e^{+4\pi i/5}\t1.
\ee 

To detect the overall phase, we initialize an ancilla qubit $|q_0\ra$ with a Hadamard gate. Re-writing the $\t1$ state in qubit basis as, 
\be
\t1 &=& |q_1 q_2 q_3 q_4\ra \\\nn
&=& \frac{1}{D}|00\ra(|10\ra + \xi^2|01\ra +\sqrt{\phi}\xi|11\ra).
\ee
results in the initial state, 
\be
|\psi_0\ra &=& |q_0\ra|q_1q_2q_3q_4\ra \\\nn
&=& \frac{1}{\sqrt{2}} (|0\ra + |1\ra)\t1.
\ee 
We then then use $|q_0\ra$ as a control qubit for the application of a controlled-$F$-move acting on $q_3$, resulting in
\be
|\psi_1\ra &=& |q_0\ra|q_1q_2q_3q_4\ra \\\nn
&=& \frac{1}{\sqrt{2}} (|0\ra + e^{4\pi i/5}|1\ra)\t1.
\ee
Thus, we use the phase kickback from twisting the $\t1$ excitation in the tadpole consisting of  qubits $q_1-q_4$ to impart a phase shift on the ancilla qubit $|q_0\ra$. We can then measure this phase shift by using the appropriate filter, which maps the state of $q_0$ to $|0\ra$. The filter in this case is the following gate,
\be
M = \begin{pmatrix}
1 &  e^{4\pi i/5} \\
e^{-4\pi i/5} & -1 \\
\end{pmatrix}.
\ee
Finally we can use $q_0$ as a control for a CNOT acting on another ancilla qubit to verify the expected phase, as in the previous cases. Since the state of the ancilla qubit $q_0$ is un-entangled from the qubits forming the tadpole, we do not need to apply a simplification circuit to these qubits. However for ease of checking the results, simplifying gates can be adopted from Fig.~\ref{fig:init} to map the $\t1$ state to $|q_1q_2q_3q_4\ra = |1110\ra$.

We end by mentioning that this procedure can easily be generalized to larger plaquettes and be used to verify the relation between braiding and twisting, Eq.~\eqref{eq:braidtwist}, in the nine-qubit lattice.

\section{Conclusions and Outlook}
\label{sec:sum}
This work was inspired by a theoretical question: what are the smallest possible lattices in which the string-net rendition of the Doubled Fibonacci model can be realized and its non-Abelian nature revealed? We answered this question by designing quantum circuits that simulate the spectrum of this model and demonstrated their fusion, twisting and braiding properties, all on minimal lattices with the fewest number of qubits possible. 

We argued that the ground state as well as a  pair of achiral $\tt$ excitations can be realized on a lattice consisting of only one qubit. We then showed that with only three qubits, we can create two pairs of $\tt$ excitations in a minimal three-plaquette lattice and demonstrate the amplitudes of their fusion channels. We further showed that by moving to a lattice with six qubits, additional fusion structure resulting from combining $\tt$ excitations can be revealed. 

We then created two pairs of chiral $\t1$ excitations in a three-plaquette tailed lattice, consisting of nine qubits, braided them and measured the relative braiding phases. Finally, we showed that a total of four qubits is enough to create a pair of chiral excitations ($\t1$ or $\1t$) in two plaquettes, twist one of them and verify the resulting phase with the aid of a fifth (ancilla) qubit. 

The Fibonacci anyon model exhibits one the simplest, yet most exotic, forms of topological order, with possible applications in active and passive fault-tolerant quantum computation. Our goal here was to find the minimal resources required to demonstrate the non-Abelian nature of Fibonacci anyons and we believe our circuits require sufficiently small quantum volume that they can be run on any modern qubit hardware with reasonable coherence times and gate fidelity, thus making this model more accessible to the wider physics community. 

The methods discussed here can  be generalized to larger lattices, where more anyons can be created and quantum gates can be carried out topologically. An important future direction would be to quantify and measure the stability of these states in the presence of noise. Another interesting direction would be the design of minimal circuits that simulate other instances of the  string-net model, resulting in other forms of topological order, especially those that require multi-level qudits.

\section*{Acknowledgment} 
We thank R. Koenig, M. Levin and A. Schotte for clarifying discussions at an earlier stage of this work, and F. Burnell for discussions and comments on an earlier draft of the manuscript. We also thank the Aspen Center for Physics
for its hospitality during the completion of part of this
work. This project was supported by the U.S. Department of Energy, Office of Science, National Quantum Information Science Research Centers, Co-design Center for Quantum Advantage~(C2QA) under contract number DE-SC0012704. LH and JP acknowledge partial support from BNL LDRD 20-022. SB and RRA were supported by the U.S. Department of Energy, Office of Science’s SULI program.  SHS is supported by EPSRC Grants
EP/S020527/1 and EP/X030881/1.

\textit{Note Added:} During the preparation of this work, we became aware of two other independently carried-out works on a related but different simulation of DFib anyons on specialized hardware~\cite{xu24} and \cite{minev24}. Besides using different methods, the present work is distinct in that it focuses on designing minimal circuits suitable for realization on any quantum platform.

\bibliography{Fib22.bib}
\end{document}

%% file: tikzmacros.tex
\usepackage{tikz}

\newcommand{\dtadpole}[9]{%
    \tikz[baseline=.1ex]{
        \foreach \i/\linecolor/\linestyle in {1/red/thick,0/black/dashed}{
            \ifnum#1=\i
                \draw [\linecolor, \linestyle] (-1.5,.5) arc (90:-90:.5); 
            \fi
            \ifnum#2=\i
                \draw [\linecolor, \linestyle] (-1.5,.5) arc (90:270:.5); 
            \fi
            \ifnum#3=\i
                \draw [\linecolor, \linestyle] (-1.5,1.5) arc (90:-90:1.5); 
            \fi
            \ifnum#4=\i
                \draw [\linecolor, \linestyle] (-1.5,1.5) arc (90:270:1.5); 
            \fi
            \ifnum#5=\i
                \draw [\linecolor, \linestyle] (-1.5,-.5) -- (-1.5,-1); 
            \fi
            \ifnum#6=\i
                \draw [\linecolor, \linestyle] (-1.5,-1) -- (-1.5,-1.5); 
            \fi
            \ifnum#7=\i
                \draw [\linecolor, \linestyle] (-1.5,.5) -- (-1.5,0); 
            \fi
            \ifnum#8=\i
                \draw [\linecolor, \linestyle] (-1.5,1.5) -- (-1.5,2); 
            \fi
            \ifnum#9=\i
                \draw [\linecolor, \linestyle] (-1.5,-1) -- (-1,-1); 
            \fi
        }
    }
}

\newcommand{\dtadpolenumbered}[9]{%
    \tikz[baseline=.1ex]{
        \foreach \i/\linecolor/\linestyle in {1/red/thick,0/black/dashed}{
            \ifnum#1=\i
                \draw [\linecolor, \linestyle] (-1.5,.5) arc (90:-90:.5); 
                \draw (-1.0,0) node[right] {0};
            \fi
            \ifnum#2=\i
                \draw [\linecolor, \linestyle] (-1.5,.5) arc (90:270:.5); 
                \draw (-2.0,0) node[left] {1};
            \fi
            \ifnum#3=\i
                \draw [\linecolor, \linestyle] (-1.5,1.5) arc (90:-90:1.5); 
                \draw (0.0,0) node[right] {2};
            \fi
            \ifnum#4=\i
                \draw [\linecolor, \linestyle] (-1.5,1.5) arc (90:270:1.5); 
                 \draw (-3.0,0) node[left] {3};
            \fi
            \ifnum#5=\i
                \draw [\linecolor, \linestyle] (-1.5,-.5) -- (-1.5,-1); 
                 \draw (-1.5,-.75) node[left] {4};
            \fi
            \ifnum#6=\i
                \draw [\linecolor, \linestyle] (-1.5,-1) -- (-1.5,-1.5); 
                \draw (-1.5,-1.25) node[right] {5};
            \fi
            \ifnum#7=\i
                \draw [\linecolor, \linestyle] (-1.5,.5) -- (-1.5,0); 
                \draw (-1.5,-.25) node[] {6};
            \fi
            \ifnum#8=\i
                \draw [\linecolor, \linestyle] (-1.5,1.5) -- (-1.5,2); 
                \draw (-1.5,1.75) node[right] {7};
            \fi
            \ifnum#9=\i
                \draw [\linecolor, \linestyle] (-1.5,-1) -- (-1,-1); 
                \draw (-1.0,-1.0) node[right] {8};
            \fi
        }
    }
}

\newcommand{\stadpole}[5]{%
    \tikz[baseline=.1ex]{
        \foreach \i/\linecolor/\linestyle in {1/red/thick,0/black/dashed}{
            \ifnum#1=\i
                \draw [\linecolor, \linestyle] (-1.5,.5) arc (90:-90:.5); 
            \fi
            \ifnum#2=\i
                \draw [\linecolor, \linestyle] (-1.5,.5) arc (90:270:.5); 
            \fi
            \ifnum#3=\i
                \draw [\linecolor, \linestyle] (-1.5,-.5) -- (-1.5,-1); 
            \fi
            \ifnum#4=\i
                \draw [\linecolor, \linestyle] (-1.5,.5) -- (-1.5,0); 
            \fi
            \ifnum#5=\i
                \draw [\linecolor, \linestyle] (-1.5,-1) -- (-1,-1); 
            \fi
        }
    }
}

\newcommand{\stadpolenumbered}[5]{%
    \tikz[baseline=.1ex]{
        \foreach \i/\linecolor/\linestyle in {1/red/thick,0/black/dashed}{
            \ifnum#1=\i
                \draw [\linecolor, \linestyle] (-1.5,.5) arc (90:-90:.5); 
                \draw (-1.0,0) node[right] {0};
            \fi
            \ifnum#2=\i
                \draw [\linecolor, \linestyle] (-1.5,.5) arc (90:270:.5); 
                \draw (-2.0,0) node[left] {1};
            \fi
            \ifnum#3=\i
                \draw [\linecolor, \linestyle] (-1.5,-.5) -- (-1.5,-1); 
                \draw (-1.5,-.75) node[left] {2};
            \fi
            \ifnum#4=\i
                \draw [\linecolor, \linestyle] (-1.5,.5) -- (-1.5,0); 
                \draw (-1.0,-1.0) node[right] {3};
            \fi
            \ifnum#5=\i
                \draw [\linecolor, \linestyle] (-1.5,-1) -- (-1,-1); 
                \draw (-1.5,-.25) node[] {4};
            \fi
        }
    }
}

\newcommand{\readoutttadpole}[9]{%
    \tikz[baseline=.1ex, scale=.75]{
        \foreach \i/\linecolor/\linestyle in {1/red/thick,0/black/dashed}{
            \ifnum#1=\i
                \draw [\linecolor, \linestyle] (-1.5,2.5) arc (90:270:2.5); 
            \fi
            \ifnum#2=\i
                \draw [\linecolor, \linestyle] (-1.5,1.5) arc (90:270:1.5); 
            \fi
            \ifnum#3=\i
                \draw [\linecolor, \linestyle] (-1.5,.5) arc (90:270:.5); 
            \fi
            \ifnum#4=\i
                \draw [\linecolor, \linestyle] (-1.5,-1) -- (-1.5,-1.5); 
            \fi
            \ifnum#5=\i
                \draw [\linecolor, \linestyle] (-1.5,2.5) arc (90:-90:2.5); 
            \fi
            \ifnum#6=\i
                \draw [\linecolor, \linestyle] (-1.5,1.5) arc (90:-90:1.5); 
            \fi
            \ifnum#7=\i
                \draw [\linecolor, \linestyle] (-1.5,1.5) -- (-1.5,2); 
            \fi
            \ifnum#8=\i
                \draw [\linecolor, \linestyle] (-1.5,.5) arc (90:-90:.5); 
            \fi 
        }
        \draw [red, thick] (-1.5,-.5) -- (-1.5,-1); 
        \draw [red, thick] (-1.5,2) -- (-1.5,2.5); 
        \draw [red, thick] (-1.5,-1) -- (-1,-1); 
        \draw [red, thick]  (-1.5,2) -- (-2,2); 
        \draw [red, thick] (-1.5,.5) -- (-1.5,0); 
        \draw [red, thick]  (-1.5,-2.5) -- (-1.5,-3); 
    }
}

\newcommand{\readoutttadpolenumbered}[9]{%
    \tikz[baseline=.1ex, scale=.75]{
        \foreach \i/\linecolor/\linestyle in {1/red/thick,0/black/dashed}{
            \ifnum#1=\i
                \draw [\linecolor, \linestyle] (-1.5,2.5) arc (90:270:2.5); 
                \draw (-4.6,0) node[right] {0};
            \fi
            \ifnum#2=\i
                \draw [\linecolor, \linestyle] (-1.5,1.5) arc (90:270:1.5); 
                \draw (-3.6,0) node[right] {1};
            \fi
            \ifnum#3=\i
                \draw [\linecolor, \linestyle] (-1.5,.5) arc (90:270:.5); 
                \draw (-2.6,0) node[right] {2};
            \fi
            \ifnum#4=\i
                \draw [\linecolor, \linestyle] (-1.5,-1) -- (-1.5,-1.5); 
                \draw (-1.4,-1.2) node[left] {3};
            \fi
            \ifnum#5=\i
                \draw [\linecolor, \linestyle] (-1.5,2.5) arc (90:-90:2.5); 
                \draw (1.0,0) node[right] {4};
            \fi
            \ifnum#6=\i
                \draw [\linecolor, \linestyle] (-1.5,1.5) arc (90:-90:1.5); 
                \draw (0.0,0) node[right] {5};
            \fi
            \ifnum#7=\i
                \draw [\linecolor, \linestyle] (-1.5,1.5) -- (-1.5,2); 
                \draw (-1.6,1.75) node[right] {7};
            \fi
            \ifnum#8=\i
                \draw [\linecolor, \linestyle] (-1.5,.5) arc (90:-90:.5); 
                \draw (-1.0,0) node[right] {9};
            \fi 
        }
        \draw [red, thick] (-1.5,-.5) -- (-1.5,-1); 
        \draw (-1.4,-.7) node[left] {6}; 
        \draw [red, thick] (-1.5,2) -- (-1.5,2.5);
        \draw (-1.6, 2.20) node[right] {8}; 
        \draw [red, thick] (-1.5,-1) -- (-1,-1); 
        \draw (-1.2,-1.0) node[right] {10}; 
        \draw [red, thick]  (-1.5,2) -- (-2,2); 
        \draw (-1.9,2.0) node[left] {11}; 
        \draw [red, thick] (-1.5,.5) -- (-1.5,0); 
        \draw (-1.5, -.2) node[] {12}; 
        \draw [red, thick]  (-1.5,-2.5) -- (-1.5,-3); 
        \draw (-1.5, -3.2) node[] {13}; 
    }
}

\newcommand{\initttadpole}[9]{%
    \tikz[baseline=.1ex, scale=.75]{
        \foreach \i/\linecolor/\linestyle in {1/red/thick,0/black/dashed}{
            \ifnum#1=\i
                \draw [\linecolor, \linestyle] (-1.5,-1.5) -- (-1.5,-2); 
            \fi
            \ifnum#2=\i
                \draw [\linecolor, \linestyle] (-1.5,.5) -- (-1.5,1); 
            \fi
            \ifnum#3=\i
                \draw [\linecolor, \linestyle] (-1.5,1.5) arc (90:-90:1.5); 
            \fi
            \ifnum#4=\i
                \draw [\linecolor, \linestyle] (-1.5,2.5) arc (90:-90:2.5); 
            \fi
            \ifnum#5=\i
                \draw [\linecolor, \linestyle] (-1.5,2.5) arc (90:270:2.5); 
            \fi
            \ifnum#6=\i
                \draw [\linecolor, \linestyle] (-1.5,1.5) arc (90:270:1.5); 
            \fi
            \ifnum#7=\i
                \draw [\linecolor, \linestyle] (-1.5,.5) arc (90:270:.5); 
            \fi
            \ifnum#8=\i
                \draw [\linecolor, \linestyle] (-1.5,.5) arc (90:-90:.5); 
            \fi 
        }
        \draw [red, thick] (-1.5,1) -- (-1.5,1.5); 
        \draw [red, thick] (-1.5,-2) -- (-1.5,-2.5); 
        \draw [red, thick] (-1.5,-.5) -- (-1.5,0); 
        \draw [red, thick] (-1.5,1) -- (-1,1); 
        \draw [red, thick]  (-1.5,-2) -- (-2,-2); 
        \draw [red, thick]  (-1.5,2.5) -- (-1.5,3); 
    }
}

\newcommand{\initttadpolenumbered}[9]{%
    \tikz[baseline=.1ex, scale=.75]{
        \foreach \i/\linecolor/\linestyle in {1/red/thick,0/black/dashed}{
            \ifnum#1=\i
                \draw [\linecolor, \linestyle] (-1.5,-1.5) -- (-1.5,-2); 
                \draw (-1.6,-1.75) node[right] {0};
            \fi
            \ifnum#2=\i
                \draw [\linecolor, \linestyle] (-1.5,.5) -- (-1.5,1); 
                \draw (-1.4,.7) node[left] {1};
            \fi
            \ifnum#3=\i
                \draw [\linecolor, \linestyle] (-1.5,1.5) arc (90:-90:1.5); 
                \draw (0.0,0) node[right] {3};
            \fi
            \ifnum#4=\i
                \draw [\linecolor, \linestyle] (-1.5,2.5) arc (90:-90:2.5); 
                \draw (1.0,0.0) node[right] {4};
            \fi
            \ifnum#5=\i
                \draw [\linecolor, \linestyle] (-1.5,2.5) arc (90:270:2.5); 
                \draw (-4.6,0) node[right] {5};
            \fi
            \ifnum#6=\i
                \draw [\linecolor, \linestyle] (-1.5,1.5) arc (90:270:1.5); 
                \draw (-3.6,0) node[right] {6};
            \fi
            \ifnum#7=\i
                \draw [\linecolor, \linestyle] (-1.5,.5) arc (90:270:.5); 
                \draw (-2.6,0) node[right] {7};
            \fi
            \ifnum#8=\i
                \draw [\linecolor, \linestyle] (-1.5,.5) arc (90:-90:.5); 
                \draw (-1.0,0) node[right] {8};
            \fi 
        }
        \draw [red, thick] (-1.5,1) -- (-1.5,1.5); 
        \draw (-1.4,1.2) node[left] {2}; 
        \draw [red, thick] (-1.5,-2) -- (-1.5,-2.5); 
        \draw (-1.6,-2.25) node[right] {9}; 
        \draw [red, thick] (-1.5,-.5) -- (-1.5,0); 
        \draw (-1.5, .2) node[] {10}; 
        \draw [red, thick] (-1.5,1) -- (-1,1); 
        \draw (-1.2,1.0) node[right] {11}; 
        \draw [red, thick]  (-1.5,-2) -- (-2,-2); 
        \draw (-1.9,-2.0) node[left] {12}; 
        \draw [red, thick]  (-1.5,2.5) -- (-1.5,3); 
        \draw (-1.5, 3.2) node[] {13}; 
    }
}

\newcommand{\tps}[4]{
  \tikz[scale=0.5]{

    \ifthenelse{\equal{#1}{1}}
    {\draw [red,thick] (-1.5,.5) arc (90:-90:.5); }
    {\draw [densely dotted] (-1.5,.5) arc (90:-90:.5); }

     \ifthenelse{\equal{#2}{1}}
    {\draw [red,thick] (-1.5,.5) arc (90:270:.5); }
    {\draw [densely dotted] (-1.5,.5) arc (90:270:.5); }

    \ifthenelse{\equal{#3}{1}}
    {\draw [red,thick] (-1.5,-.5) -- (-1.5,-1); }
    {\draw [densely dotted] (-1.5,-.5) -- (-1.5,-1); }

            \ifthenelse{\equal{#4}{1}}
            {\draw [red, thick] (-1.5,.5) -- (-1.5,0); }
            {\draw [densely dotted] (-1.5,.5) -- (-1.5,0); }
    }
}
\newcommand{\tadpole}[3]{
  \tikz[scale=0.5]{

    \ifthenelse{\equal{#1}{1}}
    {\draw [thick] (-1.5,.5) arc (90:-90:.5); }
    {\draw [densely dotted] (-1.5,.5) arc (90:-90:.5); }

     \ifthenelse{\equal{#2}{1}}
    {\draw [thick] (-1.5,.5) arc (90:270:.5); }
    {\draw [densely dotted] (-1.5,.5) arc (90:270:.5); }

    \ifthenelse{\equal{#3}{1}}
    {\draw [thick] (-1.5,-.5) -- (-1.5,-1); }
    {\draw [densely dotted] (-1.5,-.5) -- (-1.5,-1); }
    }
}

\newcommand{\gentadpole}[4]{
  \tikz[scale=0.5]{

    \ifthenelse{\equal{#1}{1}}
    {\draw [thick] (-1.5,.5) arc (90:-90:.5); }
    {\draw [densely dotted] (-1.5,.5) arc (90:-90:.5); }

     \ifthenelse{\equal{#2}{1}}
    {\draw [thick] (-1.5,.5) arc (90:270:.5); }
    {\draw [densely dotted] (-1.5,.5) arc (90:270:.5); }

    \ifthenelse{\equal{#3}{1}}
    {\draw [thick] (-1.5,-.5) -- (-1.5,-1); }
    {\draw [densely dotted] (-1.5,-.5) -- (-1.5,-1); }

            \ifthenelse{\equal{#4}{1}}
            {\draw [thick] (-1.5,.5) -- (-1.5,0); }
            {\draw [densely dotted] (-1.5,.5) -- (-1.5,0); }
    
    }
}